\documentclass[aps,pre,reprint,twocolumn,showpacs,preprintnumbers,groupedaddress,floatfix,amsmath,amssymb,nofootinbib]{revtex4-1}

\usepackage{graphicx}
\usepackage{array}
\usepackage{color}

\newcommand{\physrep}{Phys.~Rep.} % Physics Reports
\newcommand{\apjl}{Astrophys.~J.~Lett.} % The Astrophysical Journal Letters
\newcommand{\aap}{Astron.~\&~Astrophys.} % Astronomy & Astrophysics
\newcommand{\araa}{Annu.~Rev.~Astron.~Astrophys.} % Annual Review of Astronomy and Astrophysics
\newcommand{\mnras}{Mon.~Not.~R.~Astron.~Soc.} % Monthly Notices of the Royal Astronomical Society
\newcommand{\ssr}{Space Science Rev.} % Space Science Reviews
\newcommand{\njop}{New J.~of~Phys.}

\newcommand{\Rey}{\text{Re}}

\newcommand{\emag}{m}
\newcommand{\ekin}{\epsilon}

\begin{document}

\newcolumntype{C}[1]{>{\centering\arraybackslash}m{#1}}

\title{Saturation of the Turbulent Dynamo}

\author{J.~Schober}
\email[]{jschober@nordita.org}
%\thanks{}
\affiliation{Universit\"at Heidelberg, Zentrum f\"ur Astronomie, Institut f\"ur Theoretische Astrophysik, Albert-Ueberle-Strasse~2, D-69120 Heidelberg, Germany \\
and Nordita, KTH Royal Institute of Technology and Stockholm University, Roslagstullsbacken 23, 10691 Stockholm, Sweden}

\author{D.~R.~G.~Schleicher}
\affiliation{Departamento de Astronom\'ia, Facultad Ciencias F\'isicas y Matem\'aticas, Universidad de Concepci\'on, Av.~Esteban Iturra s/n Barrio Universitario, Casilla 160-C, Concepci\'on, Chile}

\author{C.~Federrath}
\affiliation{Research School of Astronomy \& Astrophysics, The Australian National University, Canberra, ACT~2611, Australia}

\author{S.~Bovino}
\affiliation{Hamburger Sternwarte, Gojenbergsweg 112, D-21029 Hamburg, Germany\\
and Institut f\"ur Astrophysik, Georg-August-Universit\"at G\"ottingen, Friedrich-Hund-Platz~1, D-37077 G\"ottingen, Germany}

\author{R.~S.~Klessen}
\affiliation{Universit\"at Heidelberg, Zentrum f\"ur Astronomie, Institut f\"ur Theoretische Astrophysik, Albert-Ueberle-Strasse~2, D-69120 Heidelberg, Germany }

\date{\today}

\begin{abstract}
The origin of strong magnetic fields in the Universe can be explained by amplifying weak seed fields via turbulent motions on small spatial scales and subsequently transporting the magnetic energy to larger scales. This process is known as the turbulent dynamo and depends on the properties of turbulence, i.e.~on the hydrodynamical Reynolds number and the compressibility of the gas, and on the magnetic diffusivity. While we know the growth rate of the magnetic energy in the linear regime, the saturation level, i.e.~the ratio of magnetic energy to turbulent kinetic energy that can be reached, is not known from analytical calculations. In this paper we present the first scale-dependent saturation model based on an effective turbulent resistivity which is determined by the turnover timescale of turbulent eddies and the magnetic energy density. The magnetic resistivity increases compared to the Spitzer value and the effective scale on which the magnetic energy spectrum is at its maximum moves to larger spatial scales. This process ends when the peak reaches a characteristic wavenumber $k_\star$ which is determined by the critical magnetic Reynolds number. The saturation level of the dynamo also depends on the type of turbulence and differs for the limits of large and small magnetic Prandtl numbers Pm. With our model we find saturation levels between 43.8\% and 1.3\% for $\mathrm{Pm}\gg1$ and between 2.43\% and 0.135\% for $\mathrm{Pm}\ll1$, where the higher values refer to incompressible turbulence and the lower ones to highly compressible turbulence.
\end{abstract}

\pacs{}
%\keywords{}

\maketitle

\section{Introduction}
\label{sec_Introduction}
%
% why B-fields are important
Magnetic fields play an important role in shaping the Universe. For understanding the formation of stars and galaxies it is thus crucial to know the strength of magnetic fields, their distribution in space, and their evolution in time. Observations in the local Universe indicate that magnetic fields are strong. In fact, the energy density of magnetic fields is often comparable to the thermal one and the one of cosmic rays, at least on spatial scales above 1 kpc \citep{StepanovEtAl2014} - a phenomenon known as energy equipartition. Moreover, magnetic fields are observed over a huge range of scales from planets \citep{Stevenson2003} and stars \citep{DonatiLandstreet2009}, to interstellar clouds \citep{Crutcher2012}, galaxies \citep{Beck2011}, and potentially the intergalactic medium \citep{NeronovVovk2010}. \\
%
% problem of amplifying seed fields -> timescales
Generation of magnetic fields is possible already in the early Universe \citep{WidrowEtAl2012}. During inflation tiny magnetic fluctuations could expand into large-scale magnetic fields. However, with flux freezing in the cosmic expansion the field strength decreases rapidly leading to very weak seed fields \citep{TurnerWidrow1988}. Likewise other generation mechanisms produce extremely weak magnetic fields, as for example field generation in cosmological phase transitions \citep{QuashnockEtAl1989,SiglEtAl1997}. Opposed to the cosmological origin of seed fields, they can also be generated in plasma processes like batteries. For instance the Biermann battery \citep{Biermann1950} can produce fields of the order of $10^{-25}-10^{-24}$ G on a comoving scale of $10$ kpc \citep{NaozNarayan2013} which is roughly 20 orders of magnitude below the $10^{-5}$ G fields observed in present-day galaxies \citep{VanEckEtAl2015}. A recent study by \citet{Schlickeiser2012} suggests that aperiodic plasma fluctuations can result in magnetic fields of the order of $10^{-16}$ G in the intergalactic medium and $10^{-10}$ G in a protogalaxy. Thus, magnetic seed fields need to be amplified very efficiently during the evolution of the Universe. \\
Let us assume that the intergalactic medium has a field strength of $B_1=10^{-20}$ G and a number density of $n_1=10^{-6}~\mathrm{cm}^{-3}$. Under the condition of flux freezing, the field strength increases only to a value of $B_2 = (n_1/n_2)^{-2/3}~B_1 = 10^{-16}$ G in spherical gravitational collapse during the formation of a galaxy with a number density of $n_2=1~\mathrm{cm}^{-3}$. Pure compression of the field lines during the collapse of a halo is thus insufficient to amplify these weak seed fields. Another possible amplification mechanism is a large-scale galactic dynamo that converts the kinetic energy from the galactic rotation into magnetic energy and simultaneously orders the fields \citep{BrandenburgSubramanian2005}. However, this process operates on long timescales comparable to the one of galactic rotation. In order to explain the observed magnetic fields in galaxies, the large-scale dynamo would require stronger seed fields than predicted by the generation mechanisms discussed above. \\
%
% kinematic turbulent dynamo -> exponential growth
There is probably only one universal mechanism by which the tiny seed fields can be amplified to the observed present-day galactic values: the \emph{small-scale or turbulent dynamo}. This type of dynamo converts kinetic energy from turbulent motions into magnetic energy by randomly stretching, twisting, and folding the field lines. As the turbulent dynamo operates initially on very small spatial scales, in fact on the diffusive scales, it is associated with very short timescales. In the kinematic dynamo phase the magnetic energy grows exponentially at large growth rates $\Gamma_\mathrm{k}$, increasing the magnetic energy on small spatial scales by many orders of magnitude. \citet{Subramanian1998} is one of the first authors who discusses the turbulent dynamo as a crucial ingredient for generating strong magnetic fields in galaxies. He argues that the latter needs to operate to provide a sufficiently strong seed field for a large-scale galactic dynamo which in turn provides an explanation for the origin of the observed fields correlated over kpc length scales. Recent semianalytical calculations \citep{SchoberEtAl2013} as well as numerical simulations \citep{LatifEtAl2013} suggest indeed that the turbulent dynamo operates efficiently already in young galaxies leading to unordered fields with a strength of the order of $10^{-5}$ G. \\
% more details on the dynamo
It has been shown that the turbulent dynamo operates under a large range of different physical conditions as long as the magnetic Reynolds number 
\begin{equation}
  \mathrm{Rm}=\frac{V L}{\eta} 
\label{eq_Rm}
\end{equation}
exceeds a critical value $\mathrm{Rm}_\mathrm{crit}$. Here $V$ is the velocity on the forcing scale $L$ of turbulence and $\eta$ is the magnetic diffusivity. Dynamo amplification is possible from very small \citep{SchekochihinEtAl2007, IskakovEtAl2007, MalyshkinBoldyrev2010, KleeorinRogachevskii2012, SchoberEtAl2012.3} to very large \citep{Subramanian1997, SchekochihinEtAl2002a, SchoberEtAl2012.1}\footnote{We refer here to the preprint paper \citet{Subramanian1997} as it includes a derivation of the Kazantsev theory for the turbulent dynamo as well as technical details on the calculation of the kinematic growth rate.} magnetic Prandtl numbers 
\begin{equation}
  \mathrm{Pm} = \frac{\nu}{\eta} = \frac{\mathrm{Rm}}{\mathrm{Re}}, 
\label{eq_Pm}
\end{equation}
where $\nu$ is the viscosity and
\begin{equation}
  \mathrm{Re}=\frac{V L}{\nu} 
\label{eq_Re}
\end{equation}
is the hydrodynamic Reynolds number. Pm describes the separation between the viscous and the resistive scale, 
\begin{equation}
  \ell_\nu = \mathrm{Re}^{-1/(1+\vartheta)}~L
\label{eq_lnu}
\end{equation}
and 
\begin{equation}
  \ell_\eta = \mathrm{Rm}^{-1/(1+\vartheta)}~L,
\label{eq_leta}
\end{equation}
and can be written as $\mathrm{Pm}=\left(\ell_\nu/\ell_\eta\right)^{1+\vartheta}$ with $\vartheta$ being the slope of the turbulence spectrum in the inertial range. The amplification process is fundamentally different for the two extreme cases of the Prandtl number. For $\mathrm{Pm}\gg1$ the dynamo operates fastest on $\ell_\nu$ leading to a growth rate of $\Gamma_\mathrm{k}\propto \mathrm{Re}^{(1-\vartheta)/(1+\vartheta)}$ \citep{SchoberEtAl2012.1}. Amplification for the case of $\mathrm{Pm}\ll1$ is most efficient on $\ell_\eta$ with a growth rate of $\Gamma_\mathrm{k}\propto \mathrm{Rm}^{(1-\vartheta)/(1+\vartheta)}$ \citep{SchoberEtAl2012.3}. Note that there is a strong dependence on the turbulence spectrum via its slope $\vartheta$ which differs for different types of turbulence. While small-scale dynamo action is most efficient in incompressible Kolmogorov turbulence with $\vartheta=1/3$ \citep{Kolmogorov1941}, it has been demonstrated that the dynamo can also operate in supersonic turbulence \citep{SchekochihinEtAl2002a, HaugenEtAl2004b, SchoberEtAl2012.1, SchoberEtAl2012.3, FederrathEtAl2014b}, i.e.~in the regime of high Mach numbers, where the gas is compressible and $\vartheta$ can reach values up to $1/2$ for Burgers-type turbulence \citep{Burgers1948}. \\
%
% short idea of our saturation model
As soon as the fields have significant back reactions on the fluid via the Lorentz force, the peak of the magnetic energy spectrum is shifted to larger spatial scales in the non-linear dynamo phase. This modification of the spectrum can be explained by an increase of the magnetic diffusivity. \citet{Subramanian1999} models an effective diffusivity which depends on the magnetic energy density and a response timescale. With the assumption that the latter is the eddy turnover time we can calculate the diffusion rate on a given spatial scale. Dynamo amplification comes to an end when the diffusion rate equals the growth rate. The magnetic energy continues to increase on larger spatial scales via stretching, twisting, and folding by larger turbulent eddies shifting the peak of the spectrum \citep{SchekochihinEtAl2002b}. Saturation of the dynamo occurs when the spectral peak reaches a maximum spatial scale which is in our model determined by an intrinsic property of dynamo action, the critical magnetic Reynolds number $\mathrm{Rm}_\mathrm{crit}$. We note that an alternative approach has been suggested in literature \citep{SchekochihinEtAl2004a} that models saturation as a result of the velocity statistics becoming anisotropic with respect to the local magnetic field.  \\
%
% results from numerical simulations
The fraction of kinetic energy that is converted into magnetic energy by the turbulent dynamo, the saturation level, has been estimated from magnetohydrodynamical (MHD) numerical simulations. For incompressible Kolmogorov-like turbulence, the saturation levels are typically 40~\% \citep{HaugenEtAl2004, FederrathEtAl2011b}. In the large Mach number regime, where the gas is highly compressible, \citet{FederrathEtAl2014b} find considerably lower values between $6\times10^{-2}$~\% and $1$~\% depending on the numerical parameters of the simulations. In this paper we aim to compare these and other numerical results from the literature with a phenomenological model for the saturation of the turbulent dynamo.\\ 
%
% paper outline
The paper is organized as follows: After a brief review on the dynamo in its different phases, the kinematic and the non-linear phase, in section \ref{sec_Amplification}, we introduce our phenomenological model of the saturation process in section \ref{sec_DynamoSaturation}. We model the behavior of the magnetic diffusivity $\eta$ which changes to an effective diffusivity $\eta_\mathrm{eff}$ in the presence of a strong magnetic field and we discuss the consequences of this for the resistive scale $\ell_\eta$, and the effective magnetic Reynolds number $\mathrm{Rm}_\mathrm{eff}=VL/\eta_\mathrm{eff}$. Dynamo amplification shows fundamental differences for the regime of large and small magnetic Prandtl numbers. Thus, we discuss the two extreme cases separately in sections \ref{subsec_LargePm} and \ref{subsec_SmallPm}. In section \ref{subsec_Simulations} we analyse the saturation of the turbulent dynamo for typical parameters of numerical simulations in order to make a direct comparison. We draw our conclusion in section \ref{sec_Conclusions}.

\section{Amplification of Random Magnetic Fields}
\label{sec_Amplification}

\begin{figure}
  \includegraphics[width=0.45\textwidth]{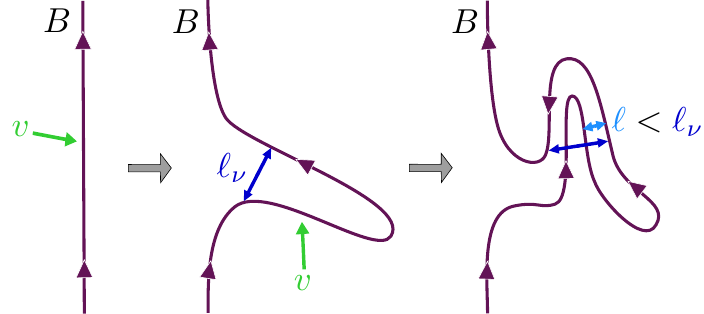}
  \caption{An illustration of the amplification of magnetic fields and the resulting transport of magnetic energy to subviscous scales. Motions of the fluid on the viscous scale $\ell_\nu$ stretch and bend the magnetic field lines to scales $\ell < \ell_\nu$. This way anti-parallel field lines are brought closer together and the magnetic energy eventually accumulates on the resistive scale $\ell_\eta$.}
\label{plot_B2SmallScales}
\end{figure}

The turbulent dynamo amplifies magnetic fields by randomly stretching, twisting, and folding the field lines in turbulent motions under the condition of flux freezing. This mechanism is most efficient on the smallest spatial scales of the system which have the shortest eddy turnover timescales. \\
%large Pm-
For large magnetic Prandtl numbers Pm, amplification is fastest on the viscous scale $\ell_\nu$ given in (\ref{eq_lnu}), where turbulent eddies are the smallest and the fastest. In fact, the magnetic energy can even be transported below the viscous scale of the system and accumulates on the resistive scale $\ell_\eta$ given in (\ref{eq_leta}). This mechanism is illustrated in figure \ref{plot_B2SmallScales}: While the turbulent eddies stretch, twist, and fold the magnetic field on a scale $\ell_\nu$, the field lines are brought closer together on spatial scales $\ell < \ell_\nu$ within multiple stretching processes. This way magnetic energy is transported down to the resistive scale $\ell_\eta$ below which it dissipates.  \\
%small Pm
For the contrary extreme of small magnetic Prandtl numbers, $\mathrm{Pm} \ll 1$, the resistive scale is above the viscous scale. In this case the dynamo operates fastest on $\ell_\eta$ and is restricted by the turnover rate of the turbulent eddies on that scale. With $\ell_\eta \gg \ell_\nu$ the typical dynamo timescales are larger for small Prandtl numbers and we expect the dynamo to be less efficient.

%%%%%%%%%%%%%%%%
%%%%%%%%%%%%%%%%

\subsection{Kinematic Dynamo Phase}
The time evolution of the magnetic field is governed by the induction equation, 
\begin{equation}
  \frac{\partial \boldsymbol{B}_\mathrm{tot}}{\partial t} = \nabla \times \left(\boldsymbol{v}_\mathrm{tot}\times\boldsymbol{B}_\mathrm{tot}\right) + \eta \mathbf{\nabla}^2 \boldsymbol{B}_\mathrm{tot},
\label{induction}
\end{equation}
which relates the total field strength $\boldsymbol{B}_\mathrm{tot}$ to the velocity field $\boldsymbol{v}_\mathrm{tot}$ and the magnetic diffusivity $\eta$. By separating the mean from the turbulent components, i.e.~$\boldsymbol{v}_\mathrm{tot} = \left\langle \boldsymbol{v}\right\rangle + \boldsymbol{v}$ and $\boldsymbol{B}_\mathrm{tot} = \left\langle \boldsymbol{B}\right\rangle + \boldsymbol{B}$, one can derive evolution equations for the mean-field and the turbulent dynamo. \\
An analytical treatment of the turbulent dynamo is possible within the framework of the Kazantsev theory \citep{Kazantsev1968, BrandenburgSubramanian2005}. A central assumption of this model is that turbulence is incompressible, where the velocity fluctuations $v$ scale with the eddy size $\ell$ as $v\propto\ell^{1/3}$. However, in astrophysical environments turbulence is usually driven by high-Mach number flows leading to a different scaling \citep{Larson1981,OssenkopfMacLow2002,ElmegreenScalo2004,ScaloElmegreen2004,MacLowKlessen2004,HeyerBrunt2004,RomanDuvalEtAl2011,KlessenGlover2014}. In this work we use
\begin{equation}
  v \propto \ell^\vartheta \propto k^{-\vartheta}
\label{vell}
\end{equation}
with $\ell\equiv2\pi/k$, where $k$ is the wavenumber\footnote{We note that we will analyse the spectra in the following mainly in Fourier space. Here we use the word ``scale'' as well for wavenumbers $k$ as for spatial lengthscales $\ell$.}, and the free parameter $\vartheta$ ranging from 1/3 \citep{Kolmogorov1941} up to 1/2 for highly compressible turbulence \citep{Burgers1948,Federrath2013}. Being interested in the statistical properties of the turbulent dynamo, we model the two-point correlation function $\left\langle v_i(\boldsymbol{r}_1,t) v_j(\boldsymbol{r}_2,s)\right\rangle$ of the fluctuating velocity field for a general type of turbulence which enters via the variable $\vartheta$. The spatial positions $\boldsymbol{r}_1$ and $\boldsymbol{r}_2$ are related by $\ell= |\boldsymbol{r}_2 -\boldsymbol{r}_1|$. If the correlation function of the magnetic field $\left\langle B_i(\boldsymbol{r}_1,t) B_j(\boldsymbol{r}_2,t)\right\rangle$ is rewritten with a separation ansatz proportional to $\psi(r)~\mathrm{exp}(2 \Gamma t)$, the growth rate $\Gamma$ is governed by the Kazantsev equation \citep{Kazantsev1968},
\begin{equation}
  -\kappa_\text{diff}(\ell)\frac{\text{d}^2\psi(\ell)}{\text{d}^2r} + U(\ell)\psi(\ell) = -\Gamma_\mathrm{k} \psi(\ell).
\label{eq_Kazantesev}
\end{equation}
The functions $\kappa_\text{diff}(r)$ and $U(r)$ depend on $\left\langle v_i(\boldsymbol{r}_1,t) v_j(\boldsymbol{r}_2,s)\right\rangle$ and $\eta$. With a model for the velocity correlation function one finds generalized expressions for $\kappa_\text{diff}(r)$ and $U(r)$. The Kazantsev equation can be solved analytically in the limits of large and small magnetic Prandtl numbers (\ref{eq_Pm}) with the result of \citep{SchoberEtAl2012.1,SchoberEtAl2012.3}
\begin{equation}
  \Gamma_\mathrm{k} = \begin{cases} 
             	              \gamma_{\mathrm{k},\mathrm{Pm}\gg1}~\dfrac{V}{L}~\mathrm{Re}^{(1-\vartheta)/(1+\vartheta)}        & ,\mathrm{Pm} \gg 1,\\ 
 \vspace{-0.4cm}\\
  	     	              \gamma_{\mathrm{k},\mathrm{Pm}\ll1}~\dfrac{V}{L}~\mathrm{Rm}^{(1-\vartheta)/(1+\vartheta)}        & ,\mathrm{Pm} \ll 1,
  	   	              \end{cases}
\label{eq_GrowthRates}
\end{equation}
with the proportionality constants 
\begin{equation}
  \gamma_{\mathrm{k},\mathrm{Pm}\gg1} =  \frac{304 \vartheta +163}{60} 
\label{eq_gammaKPmLarge}
\end{equation}
for large magnetic Prandtl numbers, and
\begin{eqnarray}
  \gamma_{\mathrm{k},\mathrm{Pm}\ll1} & = & \frac{g_1 \vartheta}{5} ~\mathrm{exp}\left(\sqrt{\frac{5}{3 g_1 \vartheta}} \pi  (\vartheta -1) -2\right)    \nonumber \\ 
 & & \times \left(\frac{\sqrt{135 g_1
   \vartheta +g_2^2}-g_2}{g_1 \vartheta }\right)^{(\vartheta -1)/(\vartheta +1)}, 
\label{eq_gammaKPmSmall}
\end{eqnarray} 
with $g_1= 56 - 103 \vartheta$ and $g_2= (79 - 157 \vartheta) \vartheta -25$ for small Prandtl numbers. The numerical values of these proportionality constants are $\gamma_{\mathrm{k},\mathrm{Pm}\gg1} =4.41-5.25$ and $\gamma_{\mathrm{k},\mathrm{Pm}\ll1}=0.0268-0.00543$ for the slopes of the turbulence spectrum $\vartheta=1/3-1/2$.
%\begin{equation}
%  \gamma_{\mathrm{k},\mathrm{Pm}\gg1} = \begin{cases} 
%             	        \frac{304 \vartheta +163}{60}      & ,\mathrm{Pm} \gg 1\\ 
%  	     	      \frac{\vartheta  a}{5} e^{\frac{\sqrt{\frac{5}{3}} \pi  (\vartheta -1)}{\sqrt{-\varthetaa}}-2} \left(\frac{\sqrt{c^2-135 \vartheta a}-c}{\vartheta a}\right)^{(\vartheta -1)(\vartheta +1)}       & ,\mathrm{Pm} \ll 1. 
%  	   	              \end{cases}
%\label{eq_GrowthRateConstants}
%\end{equation}
These solutions agree well with numerical solutions of equation (\ref{eq_Kazantesev}) \citep{BovinoEtAl2013}. \\
We note that various assumptions are made in the Kazantsev theory \citep{Kazantsev1968}. One simplification is that the turbulent velocity field is delta-correlated in time. \citet{BhatSubramanian2014} drop this assumption and derive a generalized Kazantsev equation for a finite correlation time $\tau_\mathrm{corr}$. The resulting growth rates are reduced by a factor $45/56~\tau_\mathrm{corr}$. In the classical Kazantsev model there is neither kinetic nor magnetic helicity included which would appear as an additional term in the correlation functions. \citet{MalyshkinBoldyrev2008} study the influence of kinetic helicity on small-scale dynamo amplification and find that it does not affect the fastest growing bound eigenmodes by much for incompressible turbulence.

\subsection{Non-linear Dynamo Evolution}
With the exponential growth, the fluctuating magnetic field quickly becomes strong enough to influence the velocity field via the Lorentz force. Then amplification on the dissipative scale comes to an end and the magnetic energy is shifted to larger spatial scales, typically on the eddy timescale of the current peak scale \citep{SchekochihinEtAl2002b}. The growth rate in this so-called non-linear phase is given as
\begin{equation}
  \Gamma_\mathrm{nl} = \gamma_\mathrm{nl} \frac{v(k)}{2 \pi k^{-1}},
\label{eq_GammaNL}
\end{equation}
where we determine the free parameter $\gamma_\mathrm{nl}$ self consistently in the following. The evolution of the magnetic field strength on the forcing scale is now independent of the Reynolds numbers, but still depends on the slope of the turbulence spectrum $\vartheta$. \citet{SchleicherEtAl2013} find that the field strength on the turbulent forcing scale evolves proportional to $t^{\vartheta/(1-\vartheta)}$.

%%%%%%%%%%%%%%%%%%%%%%%%%%%%%%%%%%%%%%%%%%%%%%%%%%%%%%%%%%%%%%%%%%%%%%%%%%%%%%%%%%%%%%%%%%%%%%%%%%%%
\section{The Flow of Energy over Spatial Scales and Saturation of the Turbulent Dynamo}
\label{sec_DynamoSaturation}

At some point the strength of the magnetic back reactions on the velocity field via the Lorentz force is comparable to the dynamo amplification. When this is the case on all scales of the system, the magnetic field amplification comes to an end, i.e.~the dynamo is saturated.\\
After we present a model for the magnetic diffusivity when the magnetic field approaches saturation on a given wavenumber $k$, we discuss the evolution of the magnetic energy during the different amplification stages and what determines their end. In each phase we compare the amplification rate with the dissipation rate, and finally derive the magnetic energy at saturation.

\subsection{Modification of the Magnetic Diffusivity for Strong Fields}

\citet{Subramanian1999} suggests a model for calculating the saturation energy density of the magnetic field based on describing the change of the velocity field. In fact, he introduces an effective magnetic diffusivity,
\begin{eqnarray}
  \eta_\mathrm{eff} = \eta + 2 a \frac{8 \pi}{3} \emag,
\label{eq_eta}
\end{eqnarray}
where $\eta$ is the usual microscopic resistivity, i.e.~for example the Spitzer resistivity, and the parameter $a = \tau / (4 \pi \rho)$ with $\tau$ being the response time of the system and $\rho$ being the fluid density. The second term in the effective diffusivity (\ref{eq_eta}) is proportional to the magnetic energy density $\emag$. When the dynamo reaches saturation the additional drift component dominates over the microscopic resistivity $\eta$ and we obtain from equation (\ref{eq_eta})
\begin{eqnarray}
  \eta_\mathrm{eff} \approx \frac{4}{3} \frac{\tau}{\rho} \emag.
\label{etasat}
\end{eqnarray}
The timescale $\tau$ on which the fluid reacts to the magnetic field generated by the dynamo should be comparable to the timescale of the turbulent eddies. As the eddy turnover time is different on different lengthscales, also the saturation process should be scale dependent. We model here the response time on a scale $k$ as
\begin{eqnarray}
  \tau(k)  =  \frac{2 \pi}{k v_k} 
           =  \frac{2 \pi}{V k_L^\vartheta} k^{\vartheta-1},
\label{tauell}
\end{eqnarray}
where we used equation (\ref{vell}) to find the dependence on the forcing scale and velocity, $k_L$ and $V$. \\
The rate at which the magnetic energy is dissipated can be estimated via the dissipative term in the induction equation,
\begin{equation}
  \frac{\partial \textbf{B}}{\partial t} = \eta_\mathrm{eff}(k) \mathbf{\nabla}^2 \textbf{B},
\label{induction2}
\end{equation}
where we approximate $\partial/\partial t \approx \Gamma_\mathrm{dis}$ and $\mathbf{\nabla}^2 \approx (2 \pi k^{-1})^{-2}$. The dissipation rate on a scale $k$ is then given by
\begin{equation}
  \Gamma_\mathrm{dis}(k) \approx \frac{\eta_\mathrm{eff}(k)}{(2 \pi k^{-1})^{2}}.
\label{eq_DissRate}
\end{equation}

\subsection{The Limiting Case of Large Magnetic Prandtl Numbers}
\label{subsec_LargePm}

\subsubsection{Kinematic Phase for $\mathrm{Pm}\gg1$}

In the kinematic dynamo phase the turbulent energy spectrum $E(k)$ is, per definition, not affected by the magnetic field. The total initial turbulent kinetic energy is thus determined by
\begin{eqnarray}
  \ekin_0 & = & \int_{k_L}^{k_\nu} E(k)~\mathrm{d} k   \nonumber \\
	  & = & -\frac{1}{2} \rho v(k)^2\big|_{k_L}^{k_\nu}  \nonumber \\
          & = & \frac{1}{2} \rho  V^2 \left(1-\text{Re}^{-2 \vartheta/(\vartheta +1)}\right),
\label{Ekinell}
\end{eqnarray}
where we used $v(k_L)=V$ and $v(k_\nu)=V~(k_L/k_\nu)^\vartheta$ (see equation \ref{vell}) and equation (\ref{eq_lnu}) to write this in terms of the Reynolds number. The total magnetic energy density in the kinematic phase $\emag_\mathrm{k}$ can then be determined from the integral over the magnetic energy spectrum in the kinematic phase $M_\mathrm{k}(k)$,
\begin{eqnarray}
  \emag_\mathrm{k} = \int_{k_L}^{k_{\eta}} M_\mathrm{k}(k)~\mathrm{d} k.
\label{Emagsatell}
\end{eqnarray}
A Fourier analysis of the Kazantsev theory \citep{KulsrudAnderson1992} leads to the following form of the magnetic spectrum in the kinematic phase,
\begin{eqnarray}
  M_\mathrm{k}(k) = \alpha_\mathrm{k} k^{3/2} K_0\left(k/k_{\eta}\right),
\label{Emagell}
\end{eqnarray}
where $\alpha_\mathrm{k}$ is a constant and $K_0(k/k_{\eta})$ is the modified Bessel function of the second kind. The slope which characterizes the spectrum at small wavenumbers, $k^{3/2}$, is known as the Kazantsev slope. Numerical simulations have confirmed this shape of the magnetic energy spectrum \citep{MaronCowley2001,HaugenEtAl2004}. \citet{BhatSubramanian2014} show that the $3/2$ slope of the magnetic spectrum remains even if the turbulence has a finite correlation time. In the kinematic phase the magnetic energy increases exponentially in time with a peak on the resistive scale, until: \\
\begin{equation}
  \Gamma_\mathrm{k} = \Gamma_\mathrm{dis}(\emag_\mathrm{k}, k_{\eta}).
\end{equation}
A comparison of the rates yields the total magnetic energy at the end of the kinematic phase  
\begin{equation}
  \emag_\mathrm{k} = \frac{3}{4}~\gamma_{\mathrm{k},\mathrm{Pm}\gg1}~\text{Pm}^{-1} \text{Re}^{- 2 \vartheta/(\vartheta +1)}~\rho V^2.
\label{eq_emagkinend} 
\end{equation}
For deriving equation (\ref{eq_emagkinend}) we have employed the growth rate (\ref{eq_GrowthRates}) and the dissipation rate (\ref{eq_DissRate}), where we inserted (\ref{etasat}) at $\emag_\mathrm{k}$ and $k_{\eta}$. Further we used the definition of the viscous scale (\ref{eq_lnu}) and the relation $\mathrm{Pm}=(k_\eta/k_\nu)^{1+\vartheta}$. With (\ref{eq_emagkinend}) we determine the proportionality constant $\alpha_\mathrm{k}$ in the spectrum (\ref{Emagell}) in the kinematic phase. The resulting spectrum $M_\mathrm{k}(k)$ is illustrated in figure \ref{plot_Spectra_LargePm} by the curves (a) and (b) for a hydrodynamic Reynolds number of $\Rey=10^6$ and a magnetic Prandtl number of $\mathrm{Pm}=10^3$. We present the two extreme cases of turbulence with the Kolmogorov type ($\vartheta=1/3$) in the upper panel and the Burgers type ($\vartheta=1/2$) in the lower panel. The curve (a) represents a spectrum within the kinematic phase, while curve (b) shows the spectrum at the end of the kinematic phase. Integration of curve (b) over $k$ yields the magnetic energy (\ref{eq_emagkinend}).

\subsubsection{Non-Linear Phase for $\mathrm{Pm}\gg1$}
In the non-linear dynamo phase the magnetic energy continues to increase at rate (\ref{eq_GammaNL}), while the peak of the spectrum moves to larger spatial scales, i.e.~to smaller wavenumbers $k$. It is important to note that the amplification rate is now a function of the scale $k$ on which the dynamo operates the fastest. This scale is initially the viscous scale $k_\nu$, i.e.~the scale on which the turbulent eddy timescale is the shortest. With the growth of magnetic energy, the magnetic diffusivity (\ref{eq_eta}) increases, shifting the effective resistive scale $k_{\eta,\mathrm{eff}}$ on which the magnetic spectrum peaks from $k_{\eta}$ to smaller wavenumbers. Once the peak scale of the spectrum moves to $k < k_\nu$, amplification is fastest on the current resistive scale $k_{\eta,\mathrm{eff}}$ below which the magnetic energy dissipates. \\
We thus need to model the non-linear phase in two steps. In the first step the effective resistive wavenumber is larger than the viscous wavenumber ($k_{\eta,\mathrm{eff}} > k_\nu$) and the growth rate is constant and largest on $k_\nu$. Once the peak wavenumber becomes smaller than the viscous wavenumber ($k_{\eta,\mathrm{eff}} < k_\nu$) amplification is fastest on $k_{\eta,\mathrm{eff}}$. 

\begin{figure}
  \includegraphics[width=0.5\textwidth]{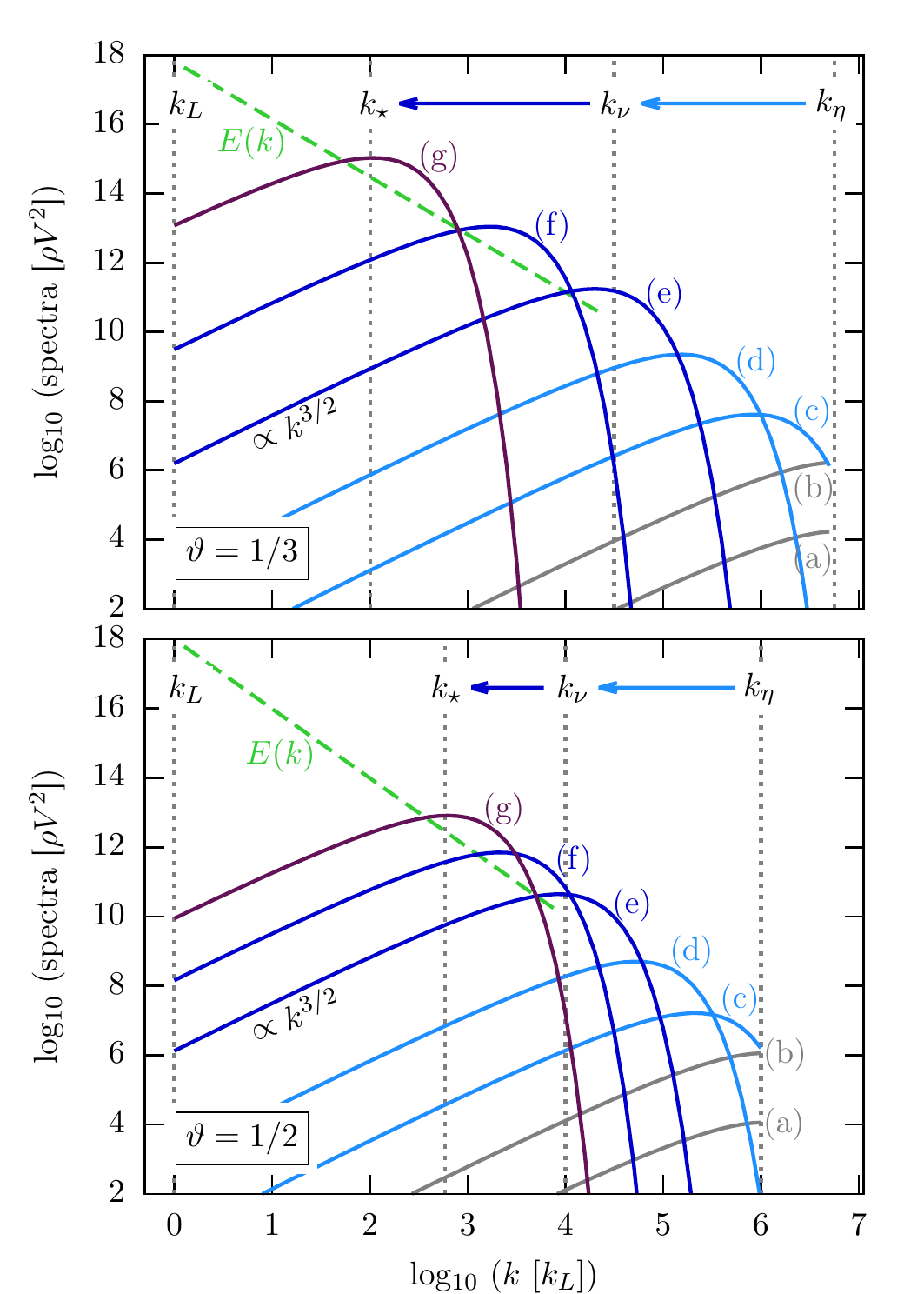}
  \caption{The spectra of turbulent kinetic and magnetic energy, $E(k)$ and $M(k)$. The input parameters for this plot are the hydrodynamic Reynolds number $\mathrm{Re}=10^6$, the magnetic Prandtl number $\mathrm{Pm}=10^3$ and the slopes of the turbulence spectrum $\vartheta=1/3$ and $\vartheta=1/2$ describing Kolmogorov and Burgers turbulence. The green dashed line indicates the (initial) kinetic energy spectrum $E(k)$. The gray lines, (a) and (b), show the magnetic energy spectrum $M_\mathrm{k}(k)$ at different times in the kinematic phase. The two light blue lines, (c) and (d), give the spectrum in the first nonlinear phase $M_\mathrm{nl,1}(k)$, while the two dark blue lines, (e) and (f), give the one in the second nonlinear phase $M_\mathrm{nl,2}(k)$. The purple line, (g), represents the magnetic spectrum at saturation $M_\mathrm{nl,2}(k_\star)$. Indicated as vertical dotted lines are the forcing scale $k_L$, the peak scale at saturation $k_\star$, the viscous scale $k_\nu$ and the initial resistive scale $k_{\eta}$.}
\label{plot_Spectra_LargePm}
\end{figure}

\paragraph{First Non-Linear Phase}
In order to have a steady increase of magnetic energy, we demand that the energy at the beginning of the first non-linear phase $\emag_\mathrm{nl,1}$ is equal to the one at the end of the kinematic phase $\emag_\mathrm{k}$. With the field amplification taking place at $k_\nu$ and dissipation taking place essentially on $k_{\eta}$ we find 
\begin{equation}
  \Gamma_\mathrm{nl}(k_\nu) = \Gamma_\mathrm{dis}(\emag_\mathrm{k}, k_{\eta})
\label{eq_ratesNL1}
\end{equation}
at the beginning of the first non-linear phase. The condition (\ref{eq_ratesNL1}) fixes the free parameter in the non-linear growth rate (\ref{eq_GammaNL}) to
\begin{equation}
  \gamma_\mathrm{nl} = \gamma_{\mathrm{k},\mathrm{Pm}\gg1}.
\label{eq_alpha}
\end{equation}
Now the peak wavelength $k_{\eta,\mathrm{eff}}$ moves to smaller $k$, while the current magnetic energy $\emag_\mathrm{nl,1}(k_{\eta,\mathrm{eff}})$ is determined by 
\begin{equation}
  \Gamma_\mathrm{nl}(k_\nu) = \Gamma_\mathrm{dis}(\emag_\mathrm{nl,1}(k_{\eta,\mathrm{eff}}), k_{\eta,\mathrm{eff}}).
\label{eq_ratesNL1_1}
\end{equation}
Solving (\ref{eq_ratesNL1_1}) for the magnetic energy and inserting the constant (\ref{eq_alpha}) results in
\begin{equation}
  \emag_\mathrm{nl,1}(k_{\eta,\mathrm{eff}}) = \frac{3}{4}~\gamma_{\mathrm{k},\mathrm{Pm}\gg1} ~ \text{Re}^{(1-\vartheta)/(\vartheta +1)} ~\left(\frac{k_\text{L}}{k_{\eta,\mathrm{eff}}}\right)^{\vartheta +1} \rho V^2.
\label{eq_emagNL1}
\end{equation}
Note that the magnetic energy increases with the peak scale $k_{\eta,\mathrm{eff}}$ moving to smaller wavenumbers $k$. The first non-linear phase comes to an end when $k_{\eta,\mathrm{eff}}$ reaches the viscous scale $k_\nu$. At this time the magnetic energy is
\begin{equation}
  \emag_\mathrm{nl,1}(k_\nu) = \frac{3}{4}~\gamma_{\mathrm{k},\mathrm{Pm}\gg1}~\text{Re}^{-2 \vartheta/(\vartheta +1)}~ \rho  V^2.
\end{equation}
We assume that the shape of the magnetic energy spectrum is similar to the one in the kinematic phase (\ref{Emagell}). With the peak of the spectrum being at the current resistive scale $k_{\eta,\mathrm{eff}}$ and the normalization changing as a function of $k_{\eta,\mathrm{eff}}$, we model the magnetic spectrum as follows:
\begin{eqnarray}
  M_\mathrm{nl,1}(k,k_{\eta,\mathrm{eff}}a) = \alpha_\mathrm{nl,1}(k_{\eta,\mathrm{eff}}) ~k^{3/2} ~K_0\left(k/k_{\eta,\mathrm{eff}}\right). \nonumber \\
\end{eqnarray}
The normalization constant $\alpha_\mathrm{nl,1}(k_{\eta,\mathrm{eff}})$ can be found using (\ref{eq_emagNL1}) and 
\begin{eqnarray}
  \emag_\mathrm{nl,1}(k_{\eta,\mathrm{eff}}) = \int_{k_L}^{k_{\eta}} M_\mathrm{nl,1}(k,k_{\eta,\mathrm{eff}})~\mathrm{d} k.
\end{eqnarray}
The resulting energy spectra are illustrated by the light-blue lines in figure \ref{plot_Spectra_LargePm}. The curve (d) shows the spectrum in the first non-linear dynamo phase at a later point in time than curve (c). One can clearly see how the peak of the spectrum moves to smaller $k$ in time and how the total magnetic energy increases from (c) to (d).
%%%%%%%%%%%%%%%%
%%%%%%%%%%%%%%%%

\paragraph{Second Non-Linear Phase}
Once the effective resistive wavenumber $k_{\eta,\mathrm{eff}}$ moves below the viscous scale $k_\nu$, magnetic field amplification becomes fastest on the current $k_{\eta,\mathrm{eff}}$. The magnetic energy $\emag_\mathrm{nl,2}$ in this phase is then determined by the condition
\begin{equation}
  \Gamma_\mathrm{nl}(k_{\eta,\mathrm{eff}}) = \Gamma_\mathrm{dis}(\emag_\mathrm{nl,2}(k_{\eta,\mathrm{eff}}), k_{\eta,\mathrm{eff}})
\label{eq_ratesNL2}
\end{equation}
which leads to
\begin{equation}
  \emag_\mathrm{nl,2}(k_{\eta,\mathrm{eff}}) = \frac{3}{4}~\gamma_{\mathrm{k},\mathrm{Pm}\gg1}~\left(\frac{k_L}{k_{\eta,\mathrm{eff}}}\right)^{2 \vartheta}~\rho V^2.
\label{eq_emagNL2}
\end{equation}
The peak of the spectrum, 
\begin{eqnarray}
  M_\mathrm{nl,2}(k,k_{\eta,\mathrm{eff}}) = \alpha_\mathrm{nl,2}(k_{\eta,\mathrm{eff}}) ~k^{3/2} ~K_0\left(k/k_{\eta,\mathrm{eff}}\right),
\end{eqnarray}
continues to move to smaller wavenumbers with the normalization constant $\alpha_\mathrm{nl,2}(k_{\eta,\mathrm{eff}})$ being determined by
\begin{eqnarray}
  \emag_\mathrm{nl,2}(k_{\eta,\mathrm{eff}}) = \int_{k_L}^{k_{\eta}} M_\mathrm{nl,2}(k,k_{\eta,\mathrm{eff}})~\mathrm{d} k,
\end{eqnarray}
when inserting the expression (\ref{eq_emagNL2}) on the left-hand side. The dark-blue lines, curves (e) and (f), in figure \ref{plot_Spectra_LargePm} show two representative spectra in the second non-linear dynamo phase. While curve (e) shows a spectrum at the very beginning of the second non-linear phase, i.e.~at a point in time when the peak has just moved below $k_\nu$, the spectrum (f) is already closer to saturation.

%%%%%%%%%%%%%%%%
\begin{table*}
   \centering
 \begin{tabular}{C{2cm} | C{2cm} | C{2cm} C{2cm} | C{2cm} C{2cm} | C{2cm} C{2cm}}
  \hline  
  \hline  
  \multicolumn{1}{c|}{$\vartheta$} & \multicolumn{1}{c|}{$\mathrm{Rm}_\mathrm{crit}$} & \multicolumn{2}{c|}{$k_\star$} & \multicolumn{2}{c|}{$\emag_\mathrm{sat}/\ekin_0$}  & \multicolumn{2}{c}{$\emag_\mathrm{sat}/(\ekin_0-\emag_\mathrm{sat})$}  \\
      ~     	& ~    		& $\mathrm{Pm}\gg1$ 		&  $\mathrm{Pm}\ll1$  		& $\mathrm{Pm}\gg1$ 		&  $\mathrm{Pm}\ll1$ & $\mathrm{Pm}\gg1$ 		&  $\mathrm{Pm}\ll1$ 	 \\
      \hline
      ~1/3            	&  $\approx 107$      	&  101~$k_L$    &  2.21~$k_L$	& 0.304	& 0.0238	& 0.438	    	& 0.0243	  \\
      ~0.35	      	&  $\approx 118$      	&  104~$k_L$	&  2.34~$k_L$ 	& 0.261	& 0.0221	& 0.352	   	& 0.0226	  \\
      ~0.37          	&  $\approx 137$      	&  110~$k_L$    &  2.51~$k_L$	& 0.212	& 0.0196	& 0.269   	& 0.0199 	\\
      ~0.38           	&  $\approx 149$      	&  114~$k_L$ 	&  2.60~$k_L$ 	& 0.190	& 0.0182	& 0.235	  	& 0.0185 	  \\
      ~0.43           	&  $\approx 227$     	&  135~$k_L$ 	&  2.78~$k_L$	& 0.108	& 0.0118	& 0.121   	& 0.0120	  \\
      ~0.47           	&  $\approx 697$     	&  260~$k_L$	&  4.15~$k_L$	& 0.0410& 0.00457	& 0.0433  	& 0.00460	  \\
      ~1/2            	&  $\approx 2718$      	&  588~$k_L$	&  6.02~$k_L$	& 0.0134& 0.00135	& 0.0136   	& 0.00136	  \\
      \hline
      \hline
  \end{tabular}
  \caption{Listed are the slopes of the turbulence spectrum $\vartheta$ for different types of turbulence (see the discussion in the text) and the corresponding critical magnetic Reynolds numbers $\mathrm{Rm}_\mathrm{crit}$ from \citep{SchoberEtAl2012.1}. For the different turbulence models we present our results for the peak scale of the spectrum $k_\star$ at saturation, and the energy ratios $\emag_\mathrm{sat}/\ekin_0$ and $\emag_\mathrm{sat}/(\ekin_0-\emag_\mathrm{sat})$. Here $\emag_\mathrm{sat}$ is the magnetic energy density at saturation and $\ekin_0$ the initial turbulent kinetic energy. For the values of the kinetic energy, we assume that we are in the limit of large hydrodynamical Reynolds numbers, where $\ekin_0 = 1/2 \rho V^2$ is constant.}
  \label{Table_Esat}
\end{table*}
%%%%%%%%%%%%%%%%

\subsubsection{Saturation for $\mathrm{Pm}\gg1$}
Saturation occurs when the peak of the magnetic energy reaches a certain scale $k_\star$ which is determined by the critical magnetic Reynolds number $\mathrm{Rm}_\mathrm{crit}$. It has been shown that a minimum Rm is needed for a turbulent dynamo to operate which can be interpreted as a minimum separation between the turbulent forcing scale and the dissipative scales. This critical magnetic Reynolds number can be found by solving the Kazantsev equation (\ref{eq_Kazantesev}) with $\Gamma=0$. A typical value for the critical magnetic Reynolds number is $\mathrm{Rm}_\mathrm{crit} \approx 100$ for Kolmogorov turbulence and increases with increasing compressibility. For Burgers turbulence $\mathrm{Rm}_\mathrm{crit} \approx 2700$ which we use for our modeling in this work. A fit formula for the critical magnetic Reynolds number is \citep{SchoberEtAl2012.1}
\begin{equation}
  \text{Rm}_\text{crit}(\vartheta) = 88\left(\text{tan}(2.7 ~\vartheta+0.2)-1\right).
\label{eq_Rmcrit}
\end{equation}
We note, however, that the values of $\mathrm{Rm}_\mathrm{crit}$ we use here are assumptions based on our previous work. There are additional numerical calculations \citep{BovinoEtAl2013} that yield a larger value of $\mathrm{Rm}_\mathrm{crit} \approx 32000$ for highly compressible turbulence. \\
With the additional drift as proposed by \citet{Subramanian1999}, an effective magnetic Reynolds number can be defined as
\begin{eqnarray}
  \mathrm{Rm}_\mathrm{eff} = \frac{V L}{\eta + 16/3 \pi ~a ~ \emag}
\end{eqnarray}
for a given magnetic energy $\emag$. Close to saturation $\mathrm{Rm}_\mathrm{eff}$ becomes
\begin{eqnarray}
  \mathrm{Rm}_\mathrm{eff} (k_{\eta,\mathrm{eff}}) \approx \frac{V L}{16/3~\pi~ a ~\emag_\mathrm{nl,2}(k_{\eta,\mathrm{eff}})},
\label{eq_RmEff}
\end{eqnarray}
where we can insert the magnetic energy in the second non-linear phase (\ref{eq_emagNL2}). It is obvious from equation (\ref{eq_RmEff}) that $\mathrm{Rm}_\mathrm{eff}$ decreases with increasing magnetic energy $\emag_\mathrm{nl,2}$ or equivalently with decreasing $k_{\eta,\mathrm{eff}}$. The decrease of $\mathrm{Rm}_\mathrm{eff}$ comes to an end as soon as it reaches $\mathrm{Rm}_\mathrm{crit}$, which is the case when $k_{\eta,\mathrm{eff}}$ reaches a certain scale $k_\star$:
\begin{eqnarray}
  \mathrm{Rm}_\mathrm{eff} (k_\star) = \mathrm{Rm}_\mathrm{crit}.
\end{eqnarray}
With equation (\ref{eq_RmEff}) and the magnetic energy in the second non-linear phase (\ref{eq_emagNL2}), we find the following value for $k_\star$:
\begin{eqnarray}
   k_\star = \left(\gamma_{\mathrm{k},\mathrm{Pm}\gg1}~\mathrm{Rm}_\mathrm{crit}\right)^{1/(\vartheta +1)} k_L.
\label{eq_kstar}
\end{eqnarray}
The numerical values of $k_\star$ are listed in table \ref{Table_Esat} for different types of turbulence. We chose the following representative turbulence slopes from literature: The two extreme cases, incompressible Kolomogorov turbulence with $\vartheta=1/3$ \citep{Kolmogorov1941} and highly compressible Burgers turbulence with $\vartheta=1/2$ \citep{Burgers1948}. If intermittency of Kolmogorov turbulence is taken into account the slope of the spectrum is $\vartheta=0.35$ \citep{SheLeveque1994}. In observations of molecular clouds  typically higher values of the order of $\vartheta=0.38$ \citep{Larson1981} and $\vartheta=0.47$ \citep{OssenkopfMacLow2002} are found. Further we list the slopes resulting from numerical simulations in driven supersonic MHD turbulence ($\vartheta=0.37$) \citep{Boldyrev2002} and from solenoidally ($\vartheta=0.43$) and compressively ($\vartheta=0.47$) driven supersonic turbulence \citep{FederrathEtAl2010}.\\
Using equation (\ref{eq_emagNL2}) the magnetic energy at saturation can be calculated as
\begin{eqnarray}
  \emag_\mathrm{sat} & = & \emag_\mathrm{nl,2} (k_\star) \nonumber \\ 
		     & = & \frac{3}{4}~\gamma_{\mathrm{k},\mathrm{Pm}\gg1} \left(\gamma_{\mathrm{k},\mathrm{Pm}\gg1} \text{Rm}_\text{crit}\right)^{-2 \vartheta/(\vartheta +1)} \rho V^2. \nonumber \\ 
\label{eq_emagsatpmlarge}
\end{eqnarray}
The spectral distribution of magnetic energy at saturation is given as
\begin{eqnarray}
   M_\mathrm{sat}(k) & = & M_\mathrm{nl,2}(k,k_\star)  \nonumber \\
		     & = & \alpha_\mathrm{nl,2}(k_\star) ~k^{3/2} ~K_0\left(k/k_\star\right),
\end{eqnarray}
where again $K_0$ denotes the Bessel function of second kind. We plot the magnetic energy spectra at saturation for the two extreme cases of turbulence, $\vartheta = 1/3$ and $\vartheta=1/2$, in figure \ref{plot_Spectra_LargePm}. One major difference between the saturation spectra (g) for the different types of turbulence is the value of $k_\star$, which is with a value of 588 $k_L$ much larger for $\vartheta=1/2$ than it is for $\vartheta=1/3$, where $k_\star=101 k_L$. Also clearly visible from figure \ref{plot_Spectra_LargePm} is that the total magnetic energy at saturation is much larger for Kolmogorov turbulence than for Burgers turbulence. \\
We define the saturation level of the turbulent dynamo as the ratio of magnetic energy at saturation $\emag_\mathrm{sat}$ to the kinetic energy at saturation. As the turbulent kinetic energy is concentrated on large spatial scales, i.e.~within the motions of the largest turbulent eddies, the kinetic energy spectrum and thus the total kinetic energy is not affected much by the growth of the magnetic field. We expect the kinetic energy at saturation to be between the two extremes of $\ekin_0$ and $\ekin_0-\emag_\mathrm{sat}$, leading to the following limiting expressions for the saturation level:
\begin{eqnarray}
  \frac{\emag_\mathrm{sat}}{\ekin_0} = \frac{3}{2}~\gamma_{\mathrm{k},\mathrm{Pm}\gg1} \left(\gamma_{\mathrm{k},\mathrm{Pm}\gg1} \text{Rm}_\text{crit}\right)^{-2 \vartheta/(\vartheta +1)}
\label{R2}
\end{eqnarray}
and
\begin{eqnarray}
   \frac{\emag_\mathrm{sat}}{\ekin_0-\emag_\mathrm{sat}} = \frac{3/2~\gamma_{\mathrm{k},\mathrm{Pm}\gg1} \left(\gamma_{\mathrm{k},\mathrm{Pm}\gg1} \text{Rm}_\text{crit}\right)^{-2 \vartheta/(\vartheta +1)}}{1 - 3/2~\gamma_{\mathrm{k},\mathrm{Pm}\gg1} \left(\gamma_{\mathrm{k},\mathrm{Pm}\gg1} \text{Rm}_\text{crit}\right)^{-2 \vartheta/(\vartheta +1)}}. \nonumber \\
\label{R1}
\end{eqnarray}
%%%%%%%%%%%%%%%%
\begin{figure}
  \includegraphics[width=0.5\textwidth]{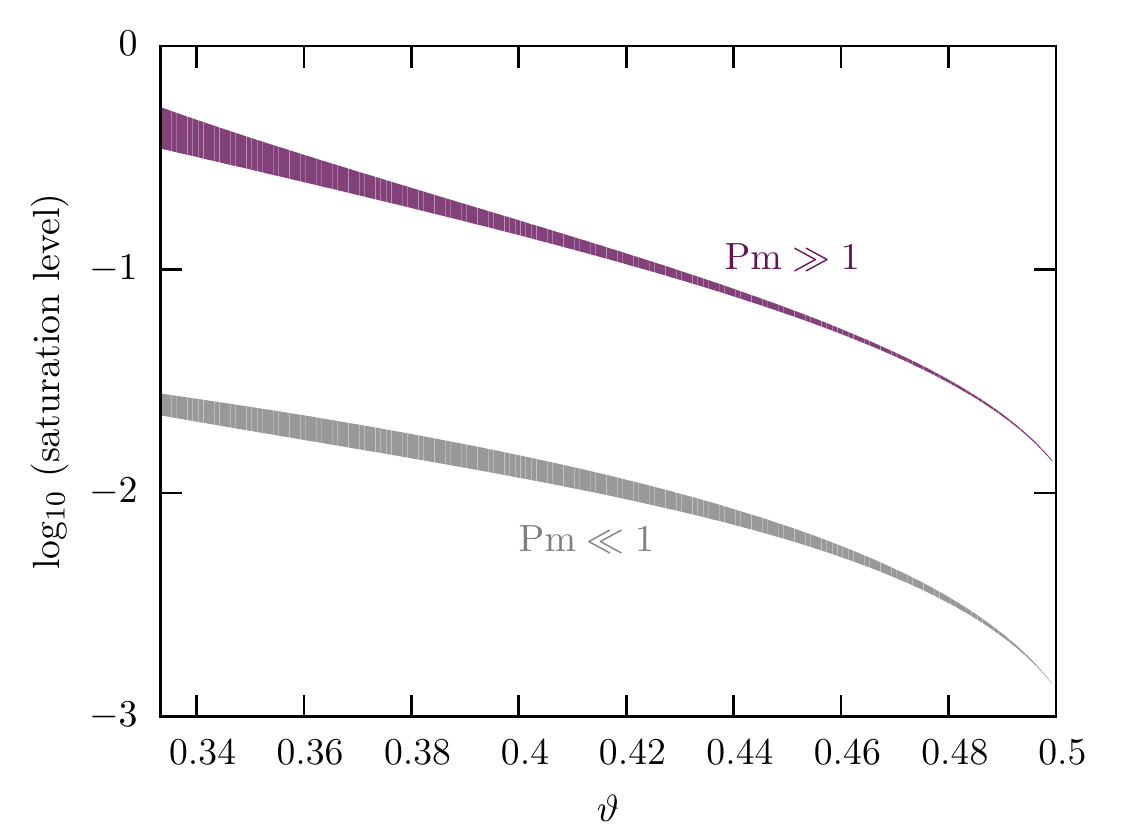}
  \caption{The saturation level of the turbulent dynamo as a function of the slope of the turbulence spectrum $\vartheta$. The upper boundaries of the bands are given by $\emag_\mathrm{sat}/(\ekin_0-\emag_\mathrm{sat})$, while the lower boundaries are determined by $\emag_\mathrm{sat}/\ekin_0$. The upper purple band represents the limit of large magnetic Prandtl numbers Pm with the saturation levels given in equations (\ref{R2}) and (\ref{R1}). The small Pm limit is illustrated by the lower gray band, where the boundaries are equations (\ref{R22}) and (\ref{R12}).}
\label{plot_SaturationLevels_slope}
\end{figure}
%%%%%%%%%%%%%%%%
The saturation level given in equation (\ref{R2}) is the ratio of the magnetic energy at saturation $\emag_\mathrm{sat}$ to the initial turbulent kinetic energy $\ekin_0$. As seen in MHD simulations \citep{FederrathEtAl2014b}, the kinetic energy spectrum does not change significantly in presence of a turbulent dynamo, leaving the total kinetic energy approximately constant. This can be achieved when the dissipation process in MHD turbulence differs from the pure hydrodynamical case and the kinetic energy is only partially converted into heat, while the remainder is transformed into magnetic energy. If the viscous dissipation is similar to the one in hydrodynamical turbulence, energy conservation requires the kinetic energy at dynamo saturation to be $\ekin_0-\emag_\mathrm{sat}$, which leads to the saturation level given in equation (\ref{R1}). \\
The two energy ratios at saturation gained from our model are listed in table \ref{Table_Esat} for different types of turbulence. Moreover, we illustrate the dependence of the saturation level on the slope of the turbulence spectrum $\vartheta$ in figure \ref{plot_SaturationLevels_slope}. Therefore we insert the fit for the critical magnetic Reynolds number (\ref{eq_Rmcrit}) and shade the area between the functions (\ref{R2}) and (\ref{R1}). The upper purple curve yields the resulting dynamo saturation level for large magnetic Prandtl numbers which decreases from approximately $30\% - 44\%$ at $\vartheta=1/3$ to roughly $1.3\%$ at $\vartheta=1/2$. Due to the small values of $\emag_\mathrm{sat}$ compared to $\ekin_0$ at very compressive turbulence, i.e.~for $\vartheta\rightarrow1/2$, the ratios $\emag_\mathrm{sat}/\ekin_0$ and $\emag_\mathrm{sat}/(\ekin_0-\emag_\mathrm{sat})$ become comparable and the band in figure \ref{plot_SaturationLevels_slope} gets very narrow.

%%%%%%%%%%%%%%%%%%%%%%%%%%%%%%%%%%%%%%%%%%%%%%%

\subsection{The Limiting Case of Small Magnetic Prandtl Numbers}
\label{subsec_SmallPm}

For small magnetic Prandtl numbers the resistive wavenumber $k_\eta$ is smaller than the viscous one $k_\nu$. Magnetic field amplification thus initially proceeds fastest on $k_\eta$ as for wavenumbers $k>k_\eta$ magnetic energy is dissipated. When the field becomes strong enough the effective resistive scale $k_{\eta,\mathrm{eff}}$ moves from $k_\eta$ to smaller $k$ until it reaches $k_\star$ and the dynamo is saturated. We discuss the evolution of the magnetic energy and its spectral distribution in the following. While the calculation is in principle similar to the case of $\mathrm{Pm} \gg 1$ we point out the main differences which occur mostly in the non-linear phase. For $\mathrm{Pm} \ll 1$ the latter is not split into two different stages as $k_\nu>k_{\eta,\mathrm{eff}}$ during the entire dynamo amplification.

\subsubsection{Kinematic Phase for $\mathrm{Pm}\ll1$}
The magnetic energy on the resistive scale $k_\eta$ increases exponentially at rate (\ref{eq_GrowthRates}) until 
\begin{equation}
  \Gamma_\mathrm{k} = \Gamma_\mathrm{dis}(\emag_\mathrm{k}, k_{\eta}).
\label{eq_RatesKin}
\end{equation}
%%%%%%%%%%%%%%%%
\begin{figure}
  \includegraphics[width=0.5\textwidth]{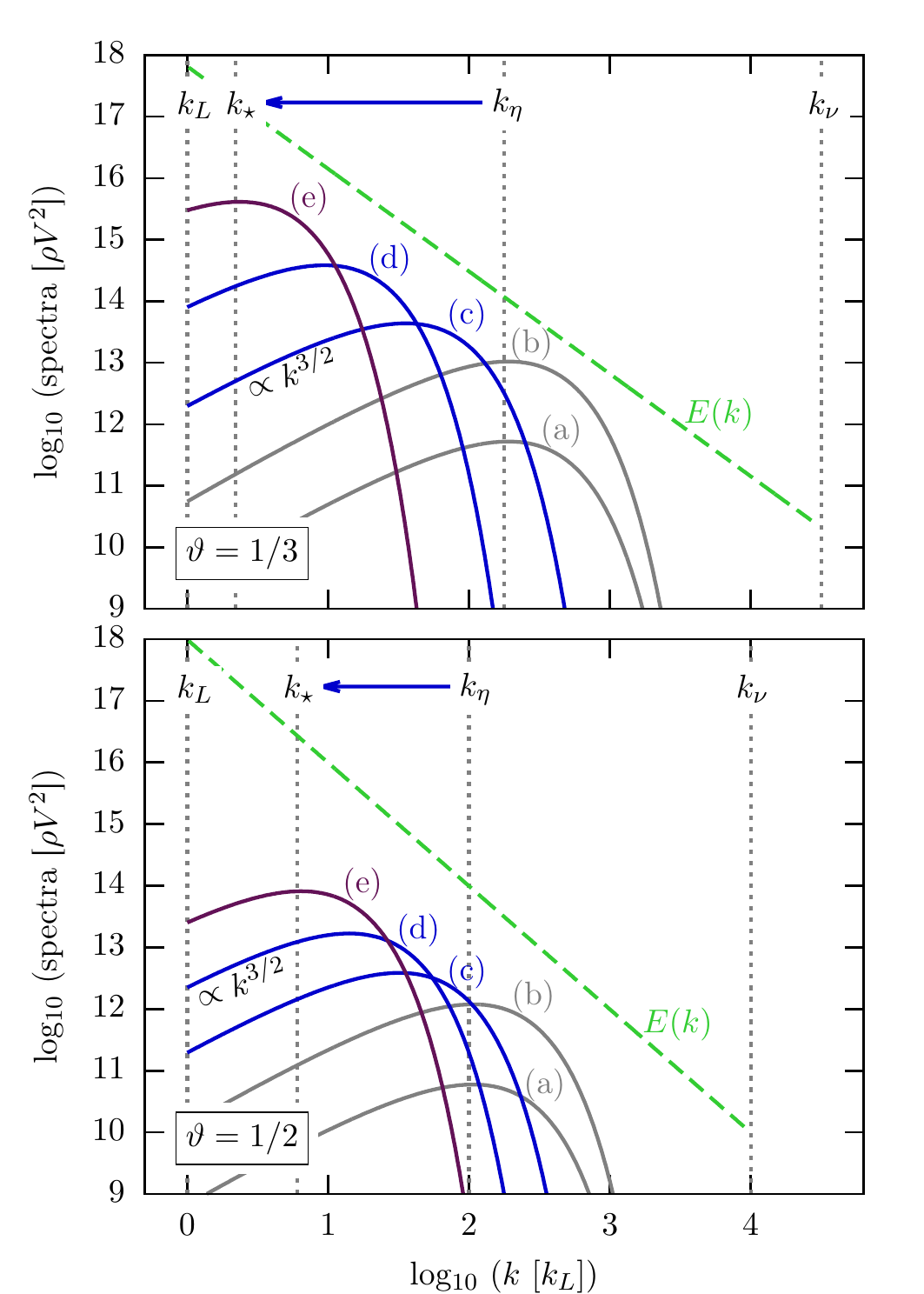}
  \caption{The spectra of turbulent kinetic and magnetic energy, $E(k)$ and $M(k)$. The input parameters for this plot are a hydrodynamic Reynolds number of $\mathrm{Re}=10^6$ and a magnetic Prandtl number of $\mathrm{Pm}=10^{-3}$. This figure is conceptually similar to figure \ref{plot_Spectra_LargePm}.}
\label{plot_Spectra_SmallPm}
\end{figure}
%%%%%%%%%%%%%%%%
Note that the growth rate (\ref{eq_GrowthRates}) scales with the magnetic Reynolds number Rm for $\mathrm{Pm} \ll 1$, while for the opposite case of $\mathrm{Pm} \gg 1$ it scales with Re. The condition (\ref{eq_RatesKin}) is fulfilled when the magnetic energy reaches
\begin{equation}
  \emag_\mathrm{k} = \frac{3}{4} \gamma_{\mathrm{k},\mathrm{Pm}\ll1} ~ \text{Rm}^{-2 \vartheta/(\vartheta +1)}~\rho  V^2.
\end{equation}
The spectral energy distribution can be determined in the same way as described in section \ref{subsec_LargePm}.

\subsubsection{Non-Linear Phase for $\mathrm{Pm}\ll1$}
The non-linear phase for the case $\mathrm{Pm}\ll1$ differs from the case of $\mathrm{Pm}\gg1$. For small magnetic Prandtl numbers the amplification proceeds initially on the resistive scale. In the transition to the non-linear phase the effective resistive scale shifts to smaller wavenumbers $k$, with amplification being always fastest on $k_{\eta,\mathrm{eff}}$. There is no separation into two different non-linear phases. \\
With the condition that the magnetic energy at the end of the kinematic phase is equal to the initial energy in the non-linear phase, we determine the proportionality constant $\gamma_\mathrm{nl}$ in the growth rate (\ref{eq_GammaNL}). With 
\begin{equation}
  \Gamma_\mathrm{nl}(k_\eta) = \Gamma_\mathrm{dis}(\emag_\mathrm{k}, k_{\eta})
\label{eq_rates_SmallPm}
\end{equation}
we find
\begin{equation}
  \gamma_\mathrm{nl} = \gamma_{\mathrm{k},\mathrm{Pm}\ll1}.
\end{equation}
We note that in equation (\ref{eq_rates_SmallPm}) we are comparing the growth rate $\Gamma_\mathrm{nl}$ on $k_\eta$ with the dissipation rate $\Gamma_\mathrm{dis}$ on $k_\eta$, while for large Pm we were comparing the growth rate on $k_\nu$ with the dissipation rate on $k_\eta$ (see equation \ref{eq_ratesNL1}). \\
For an arbitrary resistive scale $k_{\eta,\mathrm{eff}}$ the magnetic energy in the non-linear dynamo phase is determined by the condition
\begin{equation}
  \Gamma_\mathrm{nl}(k_{\eta,\mathrm{eff}}) = \Gamma_\mathrm{dis}(\emag_\mathrm{nl}(k_{\eta,\mathrm{eff}}), k_{\eta,\mathrm{eff}}).
\label{eq_EquiNL}
\end{equation}
Solution of equation (\ref{eq_EquiNL}) yields the magnetic energy 
\begin{equation}
  \emag_\mathrm{nl}(k_{\eta,\mathrm{eff}}) = \frac{3}{4} ~ \gamma_{\mathrm{k},\mathrm{Pm}\ll1} ~ \left(\frac{k_\text{L}}{k_{\eta,\mathrm{eff}}}\right)^{2 \vartheta} ~ \rho  V^2.
\end{equation}
This result for the magnetic energy is similar to the one for $\mathrm{Pm}\gg1$ given in (\ref{eq_emagNL2}) except for the different proportionality constant $\gamma_{\mathrm{k},\mathrm{Pm}\ll1}$.

\subsubsection{Saturation for $\mathrm{Pm}\ll1$}
Saturation sets in when the effective resistive scale $k_{\eta,\mathrm{eff}}$ reaches the saturation scale $k_\star$. The latter is determined via the critical magnetic Reynolds number as described in section \ref{subsec_LargePm} and is given in equation (\ref{eq_kstar}), where one needs to replace the constant $\gamma_{\mathrm{k},\mathrm{Pm}\gg1}$ by $\gamma_{\mathrm{k},\mathrm{Pm}\ll1}$:
\begin{eqnarray}
   k_\star = \left(\gamma_{\mathrm{k},\mathrm{Pm}\ll1}~\mathrm{Rm}_\mathrm{crit}\right)^{1/(\vartheta +1)} k_L.
\end{eqnarray}
The magnetic energy at saturation is then determined by
\begin{equation}
  \emag_\mathrm{nl}(k_{\star}) = \frac{3}{4} ~ \gamma_{\mathrm{k},\mathrm{Pm}\ll1} ~
   \left(\frac{k_\text{L}}{k_{\star}}\right)^{2 \vartheta} ~ \rho  V^2.
\end{equation}
Again the only difference between the saturation energy here and the case of $\mathrm{Pm}\gg1$ (\ref{eq_emagsatpmlarge}) is the proportionality constant in the growth rate (\ref{eq_GrowthRates}) which is different for small and large Prandtl numbers, see equations (\ref{eq_gammaKPmSmall}) and (\ref{eq_gammaKPmLarge}). For completeness we also give the saturation levels for the case of $\mathrm{Pm}\ll1$:
\begin{eqnarray}
  \frac{\emag_\mathrm{sat}}{\ekin_0} = \frac{3}{2}~\gamma_{\mathrm{k},\mathrm{Pm}\ll1} \left(\gamma_{\mathrm{k},\mathrm{Pm}\ll1} \text{Rm}_\text{crit}\right)^{-2 \vartheta/(\vartheta +1)}
\label{R22}
\end{eqnarray}
which is the ratio of magnetic energy at saturation $\emag_\mathrm{sat}$ to initial kinetic energy $\ekin_0$ and
\begin{eqnarray}
   \frac{\emag_\mathrm{sat}}{\ekin_0-\emag_\mathrm{sat}} = \frac{3/2~\gamma_{\mathrm{k},\mathrm{Pm}\ll1} \left(\gamma_{\mathrm{k},\mathrm{Pm}\ll1} \text{Rm}_\text{crit}\right)^{-2 \vartheta/(\vartheta +1)}}{1 - 3/2~\gamma_{\mathrm{k},\mathrm{Pm}\ll1} \left(\gamma_{\mathrm{k},\mathrm{Pm}\ll1} \text{Rm}_\text{crit}\right)^{-2 \vartheta/(\vartheta +1)}}. \nonumber \\
\label{R12}
\end{eqnarray}
We expect the dynamo to saturate at a value between (\ref{R22}) and (\ref{R12}). The evolution of the magnetic energy spectra for the case of $\mathrm{Pm}\ll1$ is presented in figure \ref{plot_Spectra_SmallPm}. The exemplary values chosen here are $\mathrm{Re}=10^6$ and $\mathrm{Pm}=10^{-3}$ and we show the two extreme cases $\vartheta=1/3$ (upper panel) and $\vartheta=1/2$ (lower panel). The curves (a) and (b) show the spectra in the kinematic phase, while (b) represents the end of the kinematic phase. Curves (c) and (d) are spectra in the non-linear phase and (e) is the spectrum at saturation. In comparison to the case of large magnetic Prandtl numbers, one clearly sees that the magnetic energy spectra are, also at saturation, much smaller. In fact, the magnetic spectra are always below the initial kinetic energy spectrum for $\mathrm{Pm}\ll1$.

%%%%%%%%%%%%%%%%%%%%%%%%%%%%%%%%%%%%%%%%%%%%%%%

\subsection{Comparison with Numerical Simulations}
\label{subsec_Simulations}

%%%%%%%%%%%%%%%%
\begin{figure}
  \includegraphics[width=0.5\textwidth]{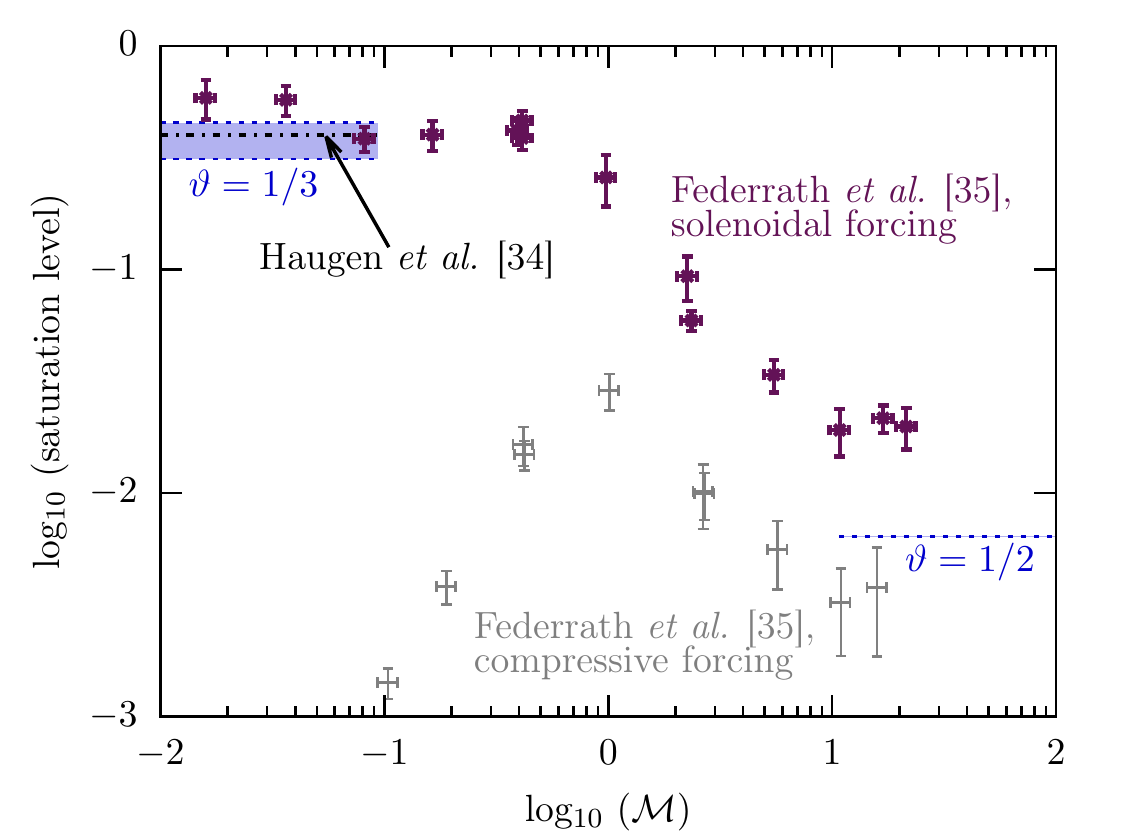}
  \caption{The saturation level of the turbulent dynamo as a function of the Mach number $\mathcal{M}$. The crosses are the results of the numerical simulations from \citet{FederrathEtAl2011b} for solenoidally driven (purple crosses) and compressively driven turbulence (gray crosses). The hydrodynamical Reynolds number of these simulations is $\mathrm{Re}\approx 1500$ and the magnetic Prandtl number is $\mathrm{Pm}\approx2$. The black dashed-dotted line at low Mach numbers indicates the saturation level in simulations of \citet{HaugenEtAl2004}. For comparison we include the predictions from our model by the blue shaded regions within the two limiting cases $\emag_\mathrm{sat}/(\ekin_0-\emag_\mathrm{sat})$ and $\emag_\mathrm{sat}/\ekin_0$ (see equations \ref{R2} and \ref{R1}) for different slopes of the turbulence spectrum $\vartheta=1/3$ and $\vartheta=1/2$.}
\label{plot_NumericalSimulations_Mach}
\end{figure}
\begin{figure}
  \includegraphics[width=0.5\textwidth]{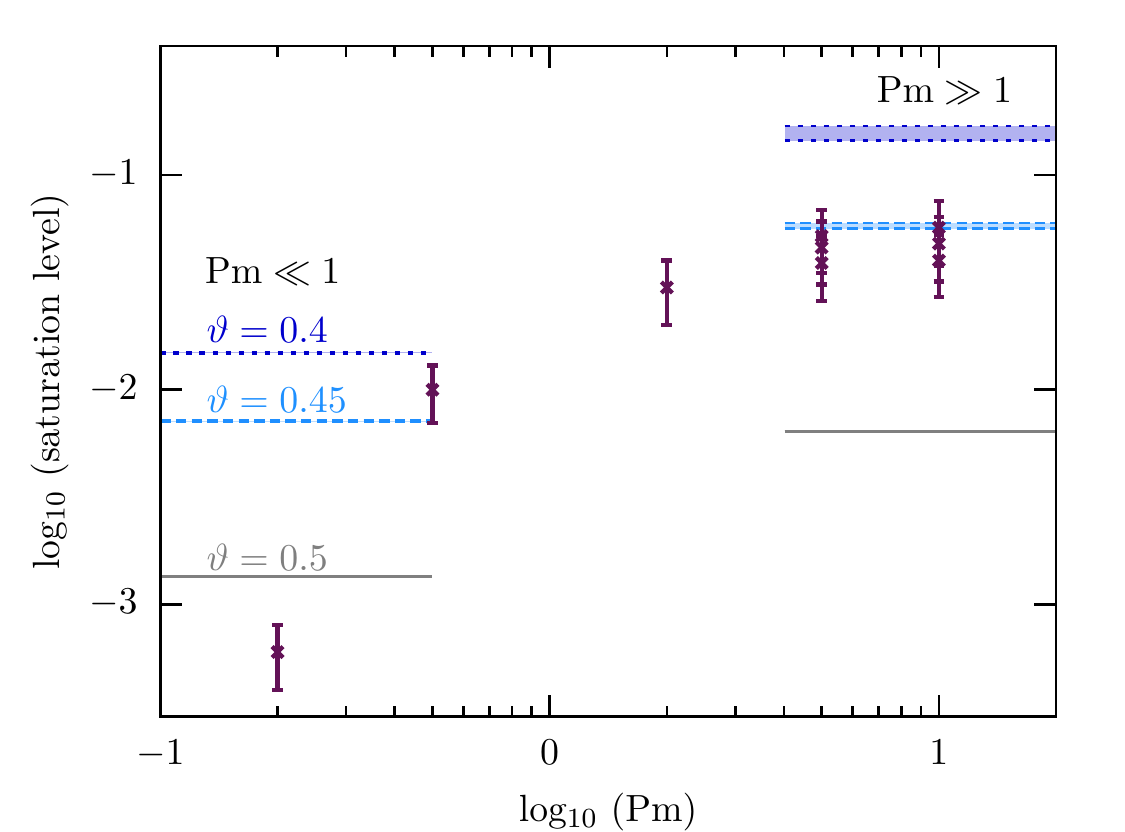}
  \caption{The saturation level of the turbulent dynamo as a function of the magnetic Prandtl number Pm. The purple crosses are the results of the numerical simulations from \citet{FederrathEtAl2014b}. In the analytical model the saturation level is independent of Pm in the limits $\mathrm{Pm}\ll 1$ and $\mathrm{Pm}\gg 1$. We shade the region between the two limiting cases $\emag_\mathrm{sat}/(\ekin_0-\emag_\mathrm{sat})$ and $\emag_\mathrm{sat}/\ekin_0$ for different slopes of the turbulence spectrum $\vartheta=0.4$ (dark blue dotted lines), $\vartheta=0.45$ (light blue dashed lines) and $\vartheta=0.5$ (gray solid lines).}
\label{plot_NumericalSimulations_Pm}
\end{figure}

 \noindent
%%%%%%%%%%%%%%%%
% intro
The turbulent dynamo has been studied intensively in numerical simulations \citep{MeneguzziEtAl1981,HaugenEtAl2004,SchekochihinEtAl2004.2,IskakovEtAl2007,FederrathEtAl2011b,FederrathEtAl2014b}. In these simulations the full set of MHD equations is solved while turbulence is initiated by a forcing term in the Navier-Stokes equation. Different types of turbulence can be simulated by applying different types of forcing, i.e.~solenoidal forcing, compressive forcing or a mixture of both, and by using different Mach numbers $\mathcal{M}$ which are defined as the ratio of the velocity dispersion to the sound speed. This way a broad range of different slopes of the turbulence spectrum $\vartheta$ can be studied. \\
% comparision with Federrath et al. 2011
The dynamo saturation levels found in the high-resolution simulations of \citet{FederrathEtAl2011b} are presented in figure \ref{plot_NumericalSimulations_Mach}. The characteristic numbers of these simulations are $\mathrm{Re}\approx 1500$ and $\mathrm{Pm}\approx2$. The ratio of magnetic energy to kinetic energy at saturation is plotted as a function of the Mach number $\mathcal{M}$ for solenoidal and compressive forcing. The sub-sonic regime of solenoidally driven turbulence is comparable to the divergence free Kolmogorov turbulence ($\vartheta=1/3$), while the supersonic regime is dominated by shocks and is better described by Burgers turbulence ($\vartheta=1/2$)\footnote{We note that for the case of $\vartheta=1/2$ and a Reynolds number of $\mathrm{Re}=1500$ the saturation scale $k_\star$ lies actually above $k_\nu$. This means that, despites being in the large Prandtl number regime, we do only have one non-linear phase. Here saturation takes place when the peak of the magnetic spectrum is still at $k>k_\nu$. This complication does not occur for $\vartheta=1/3$ and a Reynolds number of $\mathrm{Re}=1500$.}. We include in figure \ref{plot_NumericalSimulations_Mach} our predictions for the saturation levels using equations (\ref{R2}) and (\ref{R1}) as upper and lower boundaries. Note, that for $\vartheta=1/2$ the two extremes cannot be distinguished by eye in figure \ref{plot_NumericalSimulations_Mach}. The agreement between the predictions of our analytical model and the numerical simulations is very good for the case of Kolmogorov turbulence, where the saturation level is of the order of 40\%. In the supersonic regime our prediction lies with a value of roughly 0.64\% in between the saturation levels of solenoidally and compressively driven turbulence. % Haugen et al 2004
Further numerical simulations based on a different code have been performed by \citet{HaugenEtAl2004} who study non-helical turbulence without imposed large-scale fields. They use low Mach numbers to analyse the dynamo in incompressible turbulence and find a saturation level of roughly 40~\% which is in agreement with the numerical results for the saturation levels from \citet{FederrathEtAl2011b} and our analytical model for the case of incompressible turbulence. We indicate the saturation level found by \citet{HaugenEtAl2004} by the black dashed-dotted line in figure \ref{plot_NumericalSimulations_Mach}. \\
% comparision with Federrath et al. 2014
A common limitation of simulations is a relatively low separation of the dissipative scales, i.e.~$k_\nu$ and $k_\eta$, which is reflected in the magnetic Prandtl number being of the order of $\mathrm{Pm}\approx 1$. In recent high-resolution simulations, \citet{FederrathEtAl2014b} were able to analyse MHD turbulence for a larger range of magnetic Prandtl numbers from $\mathrm{Pm}\approx 0.1$ up to $\mathrm{Pm}\approx 10$. With Mach numbers up to 11, these calculations cover supersonic, highly compressible turbulence. We present the numerically found saturation levels as a function of Pm in figure \ref{plot_NumericalSimulations_Pm}. The simulations predict an increase of the saturation magnetic energy with the Prandtl number.  \\
% comparison with our model (large Pm)
We aim to compare these results with the analytical model for the saturation of the turbulent dynamo developed in the previous section. Therefore we determine the saturation levels $\emag_\mathrm{sat}/(\ekin_0)$ and $\emag_\mathrm{sat}/(\ekin_0-\emag_\mathrm{sat})$ for the two extreme cases of $\mathrm{Pm}\ll 1$ and $\mathrm{Pm}\gg 1$. We apply a hydrodynamic Reynolds number of $\mathrm{Re} =1600$ as given in \citet{FederrathEtAl2014b} and test three different slopes of the turbulence spectrum: $\vartheta=0.4$, $\vartheta=0.45$, and $\vartheta=0.5$. We expect this range of slopes for supersonic turbulence from numerical simulations with high Mach numbers \citep{FederrathEtAl2010,Federrath2013,FederrathEtAl2014b}.
% The exponent $\vartheta$ increases with the Mach number of the turbulence and approaches the Burgers value of 0.5 in the limit of very high Mach numbers. For example, \citet{FederrathEtAl2010} found $\vartheta=0.43\pm0.03$ and $\vartheta=0.47\pm0.03$ for solenoidal and compressive driving of the turbulence both with a Mach number of 5--6, while \citet{Federrath2013} obtained $\vartheta=0.48\pm0.02$ and $\vartheta=0.50\pm0.02$ for solenoidal and compressive driving, respectively, both at Mach 17 which is close to the Burgers limit. The dynamo simulations by \citet{FederrathEtAl2014b} that we use to compare with use solenoidal forcing and a Mach number of about 11, for which we expect an intermediate $\vartheta\approx0.45\pm0.05$. 
The resulting saturation levels are indicated in figure \ref{plot_NumericalSimulations_Pm}. For large magnetic Prandtl numbers we find the best agreement between the numerical results and our analytical model for a spectral slope of $\vartheta=0.45$ typical for compressive turbulence. \\
%Also the trend of a constant saturation level for large Pm seen in simulations is predicted from our analytical model. \\
% comparison with our model (small Pm)
In the contrary limit, $\mathrm{Pm}\ll 1$ and even for the regime $\mathrm{Pm}\approx 1$, \citet{FederrathEtAl2014b} find a decrease of the saturation level with decreasing Pm. In our analytical study, on the other hand, we obtain the limiting case of very small and very large Pm, without explicitly determining the behavior inbetween the two regimes. It is, however, clear that the saturation has to drop towards lower values of Pm in order to fulfill these limits. The discrepancy from the simulations at $\mathrm{Pm}<1$ may arise from the fact that for small Pm in simulations the magnetic Reynolds number of the system $\mathrm{Rm} = \mathrm{Pm}~\mathrm{Re}$ is in the vicinity of the critical magnetic Reynolds number $\mathrm{Rm}_\mathrm{crit}$. Notice, that $\mathrm{Rm}_\mathrm{crit} \approx 441$ for $\vartheta=0.45$. At $\mathrm{Pm}= 0.2$ and for $\mathrm{Re} =1600$, the magnetic Reynolds number would be $\mathrm{Rm} = 320$ and already be below $\mathrm{Rm}_\mathrm{crit}$. A direct comparison of our analytical model with the numerical simulations is thus difficult in the regime of small Pm.
%%%%%%%%%%%%%%%%
%%%%%%%%%%%%%%%%

%%%%%%%%%%%%%%%%%%%%%%%%%%%%%%%%%%%%%%%%%%%%%%%%%%%%%%%%%%%%%%%%%%%%%%%%%%%%%%%%%%%%%%%%%%%%%%%%%%%%

%%%%%%%%%%%%%%%%%%%%%%%%%%%%%%%%%%%%%%%%%%%%%%%%%%%%%%%%%%%%%%%%%%%%%%%%%%%%%%%%%%%%%%%%%%%%%%%%%%%%
\section{Conclusions}
\label{sec_Conclusions}

% what have we done
We present the first detailed scale-dependent model for the saturation of the turbulent dynamo. While it was known before that this mechanism can efficiently convert turbulent kinetic energy into magnetic energy during a phase of exponential growth, the final amplification stages and in particular the saturation levels were less clear. In this paper we present a phenomenological model for the dynamo saturation which generalizes a single-scale ansatz suggested by \citet{Subramanian1999} that includes the back reaction of the Lorentz force on the magnetic diffusivity. We assume that the term describing this back reaction is inverse proportional to the turnover time of the turbulent eddies and proportional to the magnetic energy density. By comparing the dynamo growth rate on the scale of fastest field amplification with the dissipation rate on the resistive scale we find an expression for the magnetic energy density. The latter peaks on the effective resistive spatial scale which increases in the non-linear dynamo phase shifting the magnetic energy to smaller wavenumbers. \\
%
% main results
In our semi-analytical model we distinguish the cases of large and small magnetic Prandtl numbers Pm which are fundamentally different due to the different spectral positions of the resistive and the viscous scale, $\ell_\eta$ and $\ell_\nu$. The two cases are discussed separately in sections \ref{subsec_LargePm} and \ref{subsec_SmallPm}. We find a strong dependence of the saturation levels on the slope of the initial turbulence spectrum $\vartheta$. The ratio of magnetic energy $\emag_\mathrm{sat}$ over kinetic energy at saturation lies between the extreme cases of $\emag_\mathrm{sat}/\ekin_0$ and $\emag_\mathrm{sat}/(\ekin_0-\emag_\mathrm{sat})$, where $\ekin_0$ is the initial kinetic energy. We expect that the dynamo does not effect the kinetic turbulence energy drastically and that $\ekin_0$ is maximally reduced by the saturation magnetic energy. Our main results are listed below:
\begin{itemize}
\renewcommand{\labelitemi}{$\bullet$}
\item The saturation energy levels are summarized in table \ref{Table_Esat}. For the case of $\mathrm{Pm}\gg1$ we find values between 43.8\% and 1.3\%, while the values for $\mathrm{Pm}\ll1$ are with 2.43\% to 0.135\% considerably lower. The saturation level as a function of the turbulence spectrum $\vartheta$ is plotted in figure \ref{plot_SaturationLevels_slope}.
\item The evolution of the magnetic spectra is shown in figure \ref{plot_Spectra_LargePm} for $\mathrm{Pm}\gg1$ and in figure \ref{plot_Spectra_SmallPm} for $\mathrm{Pm}\ll1$. By different representative spectra we illustrate how the magnetic energy is shifted towards smaller wavenumbers during the non-linear dynamo phase until saturation is reached.
\item For Kolmogorov turbulence ($\vartheta=1/3$) the peak of the magnetic spectrum $k_\star$ lies at $101~k_L$ for $\mathrm{Pm}\gg1$ and at $2~k_L$ for $\mathrm{Pm}\ll1$. If the turbulence is a highly compressible Burgers type ($\vartheta=1/2$) the peak scale is at $588~k_L$ for $\mathrm{Pm}\gg1$ and at $6~k_L$ for $\mathrm{Pm}\ll1$.
\end{itemize}
%
% caveats
Our model has some caveats concerning the exact shape of the energy spectra. The spectral distribution of magnetic energy (\ref{Emagell}) has explicitly been derived for the kinematic dynamo phase, when there is no back reaction from the magnetic field on the velocity field. With the Lorentz force acting on the gas, the spectral shape could in principle change. Here we make the simplest possible ansatz that the magnetic spectrum remains similar also in the non-linear dynamo phase. Another assumption is that the slope of the spectrum towards small wavenumbers is $k^{3/2}$ regardless of the slope of the turbulence spectrum. As the spectrum has been derived under the assumption of divergence-free turbulence, i.e.~Kolmogorov turbulence, it is not clear if this so-called Kazantsev slope remains for compressive turbulence. While there is first numerical evidence that the Kazantsev slope may also hold in the compressive turbulence \citep{FederrathEtAl2014b}, the detailed derivation of the magnetic energy spectrum for a dynamo in this turbulence regime remains an interesting study for the future. \\
%
% comparison with Schekochihin et al. 2004
The results from this work can be compared to the saturation model suggested by \citet{SchekochihinEtAl2004a}. These authors explain the saturation mechanism of the small-scale dynamo in incompressible and non-helical turbulence as a result of anisotropization of the velocity field with respect to the magnetic field. While they can reproduce the spectra from numerical low-Re simulations very well, they predict a scaling of the magnetic energy at saturation with $\mathrm{Re}^{-1/2}$ leading to tiny values for the extremely large Reynolds numbers in astrophysical environments. The saturation levels found with our model do not scale with the hydrodynamical Reynolds number and are moreover independent of the magnetic Prandtl number in the regimes $\mathrm{Pm}\ll1$ and $\mathrm{Pm}\gg1$. The expected field strength from these saturation levels therefore need to be considered in models of the non-linear closures of MHD turbulence, and require to consider MHD effects even in an initially weakly magnetized gas \citep{GreteEtAl2015}. \\
%
% comparison with results from shell models
An interesting alternative approach for studying the turbulent dynamo are so-called shell models. With an origin in hydrodynamic turbulence  theory these models have been extended to MHD \citep{BrandenburgEnqvistOlesen1996}. Here the Navier-Stokes equation and the induction equation are approximated by dividing them into a finite number of shells in Fourier space \citep{PlunianStepanovFrick2013}. This allows for solving the MHD equations for realistic Reynolds numbers and in particular also in the regimes $\mathrm{Pm}\ll1$ and $\mathrm{Pm}\gg1$. The exponential growth in the kinematic dynamo phase as well as the shift of the peak of the magnetic energy spectrum in the non-linear phase are well reproduced by shell models \citep{StepanovPlunian2008}. The magnetic energy spectra in the non-linear phase, however, seem to deviate from the kinematic shape given in equation (\ref{Emagell}) at high wavenumbers. We also stress that \citet{StepanovPlunian2008} report an excess of magnetic energy over kinetic energy not only for high Pm which is in agreement to our findings, but within a certain part of the inertial range also for low Pm.S \\
% in comparison to simulations
Section \ref{subsec_Simulations} of this paper is dedicated to a comparison of the predictions from our theoretical model with the results found in high-resolution numerical MHD simulations. The values found in simulations by \citet{FederrathEtAl2011b} are roughly $40$\% for Kolmogorov-type turbulence and roughly 1\% for Burgers-type turbulence. These results are very similar to the predictions of our model which are indicated by the blue bands in figure \ref{plot_NumericalSimulations_Mach}. The saturation levels as a function of the magnetic Prandtl number reported in \citet{FederrathEtAl2014b} are presented in figure \ref{plot_NumericalSimulations_Pm}. We find good agreement with our model in the large Prandtl number regime for an initial turbulence spectrum with a slope of $\vartheta=0.45$. A quantitative comparison with the simulations at $\mathrm{Pm}\ll 1$ is not possible as the magnetic Reynolds number approaches $\mathrm{Rm}_\mathrm{crit}$ leading to a rapid decrease of the saturation levels with decreasing Pm. \\
%
% what does that mean
With our phenomenological model we are able to describe the saturation of the turbulent dynamo and predict the fraction of kinetic energy that is converted into magnetic energy. In addition, we can model the magnetic spectrum to analyse how the energy is distributed over spatial scales. While numerical simulations are restricted to a regime of relatively small Reynolds and Prandtl numbers, our analytical saturation model provides for the first time estimates for MHD parameters typical for astrophysical environments.

\begin{acknowledgments}
We thank for funding through the {\em Deutsche Forschungsgemeinschaft} (DFG) in the {\em Schwer\-punkt\-programm} SPP 1573 ``Physics of the Interstellar Medium'' under grants KL 1358/14-1, SCHL 1964/1-1, SCHL 1964/1-2, and BO 4113/1-2. This work has also been financially supported by the DFG via the SFB 881 ``The Milky Way System'' in the sub-projects B1 and B2. J.~S.~and R.~S.~K.~acknowledge support from the European Research Council under the European Community's Seventh Framework Programme (FP7/2007-2013) via the ERC Advanced Grant STARLIGHT (project number 339177). C.~F.~thanks for funding provided by the Australian Research Council's Discovery Projects (grants~DP130102078 and~DP150104329).
\end{acknowledgments}


\begin{thebibliography}{60}%
\makeatletter
\providecommand \@ifxundefined [1]{%
 \@ifx{#1\undefined}
}%
\providecommand \@ifnum [1]{%
 \ifnum #1\expandafter \@firstoftwo
 \else \expandafter \@secondoftwo
 \fi
}%
\providecommand \@ifx [1]{%
 \ifx #1\expandafter \@firstoftwo
 \else \expandafter \@secondoftwo
 \fi
}%
\providecommand \natexlab [1]{#1}%
\providecommand \enquote  [1]{``#1''}%
\providecommand \bibnamefont  [1]{#1}%
\providecommand \bibfnamefont [1]{#1}%
\providecommand \citenamefont [1]{#1}%
\providecommand \href@noop [0]{\@secondoftwo}%
\providecommand \href [0]{\begingroup \@sanitize@url \@href}%
\providecommand \@href[1]{\@@startlink{#1}\@@href}%
\providecommand \@@href[1]{\endgroup#1\@@endlink}%
\providecommand \@sanitize@url [0]{\catcode `\\12\catcode `\$12\catcode
  `\&12\catcode `\#12\catcode `\^12\catcode `\_12\catcode `\%12\relax}%
\providecommand \@@startlink[1]{}%
\providecommand \@@endlink[0]{}%
\providecommand \url  [0]{\begingroup\@sanitize@url \@url }%
\providecommand \@url [1]{\endgroup\@href {#1}{\urlprefix }}%
\providecommand \urlprefix  [0]{URL }%
\providecommand \Eprint [0]{\href }%
\providecommand \doibase [0]{http://dx.doi.org/}%
\providecommand \selectlanguage [0]{\@gobble}%
\providecommand \bibinfo  [0]{\@secondoftwo}%
\providecommand \bibfield  [0]{\@secondoftwo}%
\providecommand \translation [1]{[#1]}%
\providecommand \BibitemOpen [0]{}%
\providecommand \bibitemStop [0]{}%
\providecommand \bibitemNoStop [0]{.\EOS\space}%
\providecommand \EOS [0]{\spacefactor3000\relax}%
\providecommand \BibitemShut  [1]{\csname bibitem#1\endcsname}%
\let\auto@bib@innerbib\@empty
%</preamble>
\bibitem [{\citenamefont {{Stepanov}}\ \emph {et~al.}(2014)\citenamefont
  {{Stepanov}}, \citenamefont {{Shukurov}}, \citenamefont {{Fletcher}},
  \citenamefont {{Beck}}, \citenamefont {{La Porta}},\ and\ \citenamefont
  {{Tabatabaei}}}]{StepanovEtAl2014}%
  \BibitemOpen
  \bibfield  {author} {\bibinfo {author} {\bibfnamefont {R.}~\bibnamefont
  {{Stepanov}}}, \bibinfo {author} {\bibfnamefont {A.}~\bibnamefont
  {{Shukurov}}}, \bibinfo {author} {\bibfnamefont {A.}~\bibnamefont
  {{Fletcher}}}, \bibinfo {author} {\bibfnamefont {R.}~\bibnamefont {{Beck}}},
  \bibinfo {author} {\bibfnamefont {L.}~\bibnamefont {{La Porta}}}, \ and\
  \bibinfo {author} {\bibfnamefont {F.}~\bibnamefont {{Tabatabaei}}},\ }\href
  {\doibase 10.1093/mnras/stt2044} {\bibfield  {journal} {\bibinfo  {journal}
  {\mnras}\ }\textbf {\bibinfo {volume} {437}},\ \bibinfo {pages} {2201}
  (\bibinfo {year} {2014})}\BibitemShut {NoStop}%
\bibitem [{\citenamefont {{Stevenson}}(2003)}]{Stevenson2003}%
  \BibitemOpen
  \bibfield  {author} {\bibinfo {author} {\bibfnamefont {D.~J.}\ \bibnamefont
  {{Stevenson}}},\ }\href {\doibase 10.1016/S0012-821X(02)01126-3} {\bibfield
  {journal} {\bibinfo  {journal} {Earth and Planet.~Sci.~Let.~}\ }\textbf
  {\bibinfo {volume} {208}},\ \bibinfo {pages} {1} (\bibinfo {year}
  {2003})}\BibitemShut {NoStop}%
\bibitem [{\citenamefont {{Donati}}\ and\ \citenamefont
  {{Landstreet}}(2009)}]{DonatiLandstreet2009}%
  \BibitemOpen
  \bibfield  {author} {\bibinfo {author} {\bibfnamefont {J.-F.}\ \bibnamefont
  {{Donati}}}\ and\ \bibinfo {author} {\bibfnamefont {J.~D.}\ \bibnamefont
  {{Landstreet}}},\ }\href {\doibase 10.1146/annurev-astro-082708-101833}
  {\bibfield  {journal} {\bibinfo  {journal} {\araa}\ }\textbf {\bibinfo
  {volume} {47}},\ \bibinfo {pages} {333} (\bibinfo {year} {2009})}\BibitemShut
  {NoStop}%
\bibitem [{\citenamefont {{Crutcher}}(2012)}]{Crutcher2012}%
  \BibitemOpen
  \bibfield  {author} {\bibinfo {author} {\bibfnamefont {R.~M.}\ \bibnamefont
  {{Crutcher}}},\ }\href {\doibase 10.1146/annurev-astro-081811-125514}
  {\bibfield  {journal} {\bibinfo  {journal} {\araa}\ }\textbf {\bibinfo
  {volume} {50}},\ \bibinfo {pages} {29} (\bibinfo {year} {2012})}\BibitemShut
  {NoStop}%
\bibitem [{\citenamefont {{Beck}}(2011)}]{Beck2011}%
  \BibitemOpen
  \bibfield  {author} {\bibinfo {author} {\bibfnamefont {R.}~\bibnamefont
  {{Beck}}},\ }\href {\doibase 10.1007/s11214-011-9782-z} {\bibfield  {journal}
  {\bibinfo  {journal} {\ssr}\ ,\ \bibinfo {pages} {135}} (\bibinfo {year}
  {2011})}\BibitemShut {NoStop}%
\bibitem [{\citenamefont {{Neronov}}\ and\ \citenamefont
  {{Vovk}}(2010)}]{NeronovVovk2010}%
  \BibitemOpen
  \bibfield  {author} {\bibinfo {author} {\bibfnamefont {A.}~\bibnamefont
  {{Neronov}}}\ and\ \bibinfo {author} {\bibfnamefont {I.}~\bibnamefont
  {{Vovk}}},\ }\href {\doibase 10.1126/science.1184192} {\bibfield  {journal}
  {\bibinfo  {journal} {Science}\ }\textbf {\bibinfo {volume} {328}},\ \bibinfo
  {pages} {73} (\bibinfo {year} {2010})}\BibitemShut {NoStop}%
\bibitem [{\citenamefont {{Widrow}}\ \emph {et~al.}(2012)\citenamefont
  {{Widrow}}, \citenamefont {{Ryu}}, \citenamefont {{Schleicher}},
  \citenamefont {{Subramanian}}, \citenamefont {{Tsagas}},\ and\ \citenamefont
  {{Treumann}}}]{WidrowEtAl2012}%
  \BibitemOpen
  \bibfield  {author} {\bibinfo {author} {\bibfnamefont {L.~M.}\ \bibnamefont
  {{Widrow}}}, \bibinfo {author} {\bibfnamefont {D.}~\bibnamefont {{Ryu}}},
  \bibinfo {author} {\bibfnamefont {D.~R.~G.}\ \bibnamefont {{Schleicher}}},
  \bibinfo {author} {\bibfnamefont {K.}~\bibnamefont {{Subramanian}}}, \bibinfo
  {author} {\bibfnamefont {C.~G.}\ \bibnamefont {{Tsagas}}}, \ and\ \bibinfo
  {author} {\bibfnamefont {R.~A.}\ \bibnamefont {{Treumann}}},\ }\href
  {\doibase 10.1007/s11214-011-9833-5} {\bibfield  {journal} {\bibinfo
  {journal} {\ssr}\ }\textbf {\bibinfo {volume} {166}},\ \bibinfo {pages} {37}
  (\bibinfo {year} {2012})}\BibitemShut {NoStop}%
\bibitem [{\citenamefont {Turner}\ and\ \citenamefont
  {Widrow}(1988)}]{TurnerWidrow1988}%
  \BibitemOpen
  \bibfield  {author} {\bibinfo {author} {\bibfnamefont {M.~S.}\ \bibnamefont
  {Turner}}\ and\ \bibinfo {author} {\bibfnamefont {L.~M.}\ \bibnamefont
  {Widrow}},\ }\href {\doibase 10.1103/PhysRevD.37.2743} {\bibfield  {journal}
  {\bibinfo  {journal} {Phys.~Rev.~D}\ }\textbf {\bibinfo {volume} {37}},\
  \bibinfo {pages} {2743} (\bibinfo {year} {1988})}\BibitemShut {NoStop}%
\bibitem [{\citenamefont {{Quashnock}}\ \emph {et~al.}(1989)\citenamefont
  {{Quashnock}}, \citenamefont {{Loeb}},\ and\ \citenamefont
  {{Spergel}}}]{QuashnockEtAl1989}%
  \BibitemOpen
  \bibfield  {author} {\bibinfo {author} {\bibfnamefont {J.~M.}\ \bibnamefont
  {{Quashnock}}}, \bibinfo {author} {\bibfnamefont {A.}~\bibnamefont {{Loeb}}},
  \ and\ \bibinfo {author} {\bibfnamefont {D.~N.}\ \bibnamefont {{Spergel}}},\
  }\href {\doibase 10.1086/185528} {\bibfield  {journal} {\bibinfo  {journal}
  {\apjl}\ }\textbf {\bibinfo {volume} {344}},\ \bibinfo {pages} {L49}
  (\bibinfo {year} {1989})}\BibitemShut {NoStop}%
\bibitem [{\citenamefont {{Sigl}}\ \emph {et~al.}(1997)\citenamefont {{Sigl}},
  \citenamefont {{Olinto}},\ and\ \citenamefont {{Jedamzik}}}]{SiglEtAl1997}%
  \BibitemOpen
  \bibfield  {author} {\bibinfo {author} {\bibfnamefont {G.}~\bibnamefont
  {{Sigl}}}, \bibinfo {author} {\bibfnamefont {A.~V.}\ \bibnamefont
  {{Olinto}}}, \ and\ \bibinfo {author} {\bibfnamefont {K.}~\bibnamefont
  {{Jedamzik}}},\ }\href {\doibase 10.1103/PhysRevD.55.4582} {\bibfield
  {journal} {\bibinfo  {journal} {\prd}\ }\textbf {\bibinfo {volume} {55}},\
  \bibinfo {pages} {4582} (\bibinfo {year} {1997})}\BibitemShut {NoStop}%
\bibitem [{\citenamefont {{Biermann}}(1950)}]{Biermann1950}%
  \BibitemOpen
  \bibfield  {author} {\bibinfo {author} {\bibfnamefont {L.}~\bibnamefont
  {{Biermann}}},\ }\href@noop {} {\bibfield  {journal} {\bibinfo  {journal}
  {Zeitschrift Naturforschung Teil A}\ }\textbf {\bibinfo {volume} {5}},\
  \bibinfo {pages} {65} (\bibinfo {year} {1950})}\BibitemShut {NoStop}%
\bibitem [{\citenamefont {{Naoz}}\ and\ \citenamefont
  {{Narayan}}(2013)}]{NaozNarayan2013}%
  \BibitemOpen
  \bibfield  {author} {\bibinfo {author} {\bibfnamefont {S.}~\bibnamefont
  {{Naoz}}}\ and\ \bibinfo {author} {\bibfnamefont {R.}~\bibnamefont
  {{Narayan}}},\ }\href {\doibase 10.1103/PhysRevLett.111.051303} {\bibfield
  {journal} {\bibinfo  {journal} {Physical Review Letters}\ }\textbf {\bibinfo
  {volume} {111}},\ \bibinfo {eid} {051303} (\bibinfo {year}
  {2013})}\BibitemShut {NoStop}%
\bibitem [{\citenamefont {{Van Eck}}\ \emph {et~al.}(2015)\citenamefont {{Van
  Eck}}, \citenamefont {{Brown}}, \citenamefont {{Shukurov}},\ and\
  \citenamefont {{Fletcher}}}]{VanEckEtAl2015}%
  \BibitemOpen
  \bibfield  {author} {\bibinfo {author} {\bibfnamefont {C.~L.}\ \bibnamefont
  {{Van Eck}}}, \bibinfo {author} {\bibfnamefont {J.~C.}\ \bibnamefont
  {{Brown}}}, \bibinfo {author} {\bibfnamefont {A.}~\bibnamefont {{Shukurov}}},
  \ and\ \bibinfo {author} {\bibfnamefont {A.}~\bibnamefont {{Fletcher}}},\
  }\href {\doibase 10.1088/0004-637X/799/1/35} {\bibfield  {journal} {\bibinfo
  {journal} {\apj}\ }\textbf {\bibinfo {volume} {799}},\ \bibinfo {eid} {35}
  (\bibinfo {year} {2015})},\ \Eprint {http://arxiv.org/abs/1411.1386}
  {arXiv:1411.1386} \BibitemShut {NoStop}%
\bibitem [{\citenamefont {{Schlickeiser}}(2012)}]{Schlickeiser2012}%
  \BibitemOpen
  \bibfield  {author} {\bibinfo {author} {\bibfnamefont {R.}~\bibnamefont
  {{Schlickeiser}}},\ }\href {\doibase 10.1103/PhysRevLett.109.261101}
  {\bibfield  {journal} {\bibinfo  {journal} {\prl}\ }\textbf {\bibinfo
  {volume} {109}},\ \bibinfo {eid} {261101} (\bibinfo {year}
  {2012})}\BibitemShut {NoStop}%
\bibitem [{\citenamefont {{Brandenburg}}\ and\ \citenamefont
  {{Subramanian}}(2005)}]{BrandenburgSubramanian2005}%
  \BibitemOpen
  \bibfield  {author} {\bibinfo {author} {\bibfnamefont {A.}~\bibnamefont
  {{Brandenburg}}}\ and\ \bibinfo {author} {\bibfnamefont {K.}~\bibnamefont
  {{Subramanian}}},\ }\href {\doibase 10.1016/j.physrep.2005.06.005} {\bibfield
   {journal} {\bibinfo  {journal} {\physrep}\ }\textbf {\bibinfo {volume}
  {417}},\ \bibinfo {pages} {1} (\bibinfo {year} {2005})}\BibitemShut {NoStop}%
\bibitem [{\citenamefont {{Subramanian}}(1998)}]{Subramanian1998}%
  \BibitemOpen
  \bibfield  {author} {\bibinfo {author} {\bibfnamefont {K.}~\bibnamefont
  {{Subramanian}}},\ }\href {\doibase 10.1046/j.1365-8711.1998.01284.x}
  {\bibfield  {journal} {\bibinfo  {journal} {\mnras}\ }\textbf {\bibinfo
  {volume} {294}},\ \bibinfo {pages} {718} (\bibinfo {year}
  {1998})}\BibitemShut {NoStop}%
\bibitem [{\citenamefont {{Schober}}\ \emph {et~al.}(2013)\citenamefont
  {{Schober}}, \citenamefont {{Schleicher}},\ and\ \citenamefont
  {{Klessen}}}]{SchoberEtAl2013}%
  \BibitemOpen
  \bibfield  {author} {\bibinfo {author} {\bibfnamefont {J.}~\bibnamefont
  {{Schober}}}, \bibinfo {author} {\bibfnamefont {D.~R.~G.}\ \bibnamefont
  {{Schleicher}}}, \ and\ \bibinfo {author} {\bibfnamefont {R.~S.}\
  \bibnamefont {{Klessen}}},\ }\href {\doibase 10.1051/0004-6361/201322185}
  {\bibfield  {journal} {\bibinfo  {journal} {\aap}\ }\textbf {\bibinfo
  {volume} {560}},\ \bibinfo {eid} {A87} (\bibinfo {year} {2013})}\BibitemShut
  {NoStop}%
\bibitem [{\citenamefont {{Latif}}\ \emph {et~al.}(2013)\citenamefont
  {{Latif}}, \citenamefont {{Schleicher}}, \citenamefont {{Schmidt}},\ and\
  \citenamefont {{Niemeyer}}}]{LatifEtAl2013}%
  \BibitemOpen
  \bibfield  {author} {\bibinfo {author} {\bibfnamefont {M.~A.}\ \bibnamefont
  {{Latif}}}, \bibinfo {author} {\bibfnamefont {D.~R.~G.}\ \bibnamefont
  {{Schleicher}}}, \bibinfo {author} {\bibfnamefont {W.}~\bibnamefont
  {{Schmidt}}}, \ and\ \bibinfo {author} {\bibfnamefont {J.}~\bibnamefont
  {{Niemeyer}}},\ }\href {\doibase 10.1093/mnras/stt503} {\bibfield  {journal}
  {\bibinfo  {journal} {\mnras}\ } (\bibinfo {year} {2013}),\
  10.1093/mnras/stt503}\BibitemShut {NoStop}%
\bibitem [{\citenamefont {{Schekochihin}}\ \emph {et~al.}(2007)\citenamefont
  {{Schekochihin}}, \citenamefont {{Iskakov}}, \citenamefont {{Cowley}},
  \citenamefont {{McWilliams}}, \citenamefont {{Proctor}},\ and\ \citenamefont
  {{Yousef}}}]{SchekochihinEtAl2007}%
  \BibitemOpen
  \bibfield  {author} {\bibinfo {author} {\bibfnamefont {A.~A.}\ \bibnamefont
  {{Schekochihin}}}, \bibinfo {author} {\bibfnamefont {A.~B.}\ \bibnamefont
  {{Iskakov}}}, \bibinfo {author} {\bibfnamefont {S.~C.}\ \bibnamefont
  {{Cowley}}}, \bibinfo {author} {\bibfnamefont {J.~C.}\ \bibnamefont
  {{McWilliams}}}, \bibinfo {author} {\bibfnamefont {M.~R.~E.}\ \bibnamefont
  {{Proctor}}}, \ and\ \bibinfo {author} {\bibfnamefont {T.~A.}\ \bibnamefont
  {{Yousef}}},\ }\href {\doibase 10.1088/1367-2630/9/8/300} {\bibfield
  {journal} {\bibinfo  {journal} {\njop}\ }\textbf {\bibinfo {volume} {9}},\
  \bibinfo {pages} {300} (\bibinfo {year} {2007})}\BibitemShut {NoStop}%
\bibitem [{\citenamefont {{Iskakov}}\ \emph {et~al.}(2007)\citenamefont
  {{Iskakov}}, \citenamefont {{Schekochihin}}, \citenamefont {{Cowley}},
  \citenamefont {{McWilliams}},\ and\ \citenamefont
  {{Proctor}}}]{IskakovEtAl2007}%
  \BibitemOpen
  \bibfield  {author} {\bibinfo {author} {\bibfnamefont {A.~B.}\ \bibnamefont
  {{Iskakov}}}, \bibinfo {author} {\bibfnamefont {A.~A.}\ \bibnamefont
  {{Schekochihin}}}, \bibinfo {author} {\bibfnamefont {S.~C.}\ \bibnamefont
  {{Cowley}}}, \bibinfo {author} {\bibfnamefont {J.~C.}\ \bibnamefont
  {{McWilliams}}}, \ and\ \bibinfo {author} {\bibfnamefont {M.~R.~E.}\
  \bibnamefont {{Proctor}}},\ }\href {\doibase 10.1103/PhysRevLett.98.208501}
  {\bibfield  {journal} {\bibinfo  {journal} {\prl}\ }\textbf {\bibinfo
  {volume} {98}},\ \bibinfo {eid} {208501} (\bibinfo {year}
  {2007})}\BibitemShut {NoStop}%
\bibitem [{\citenamefont {Malyshkin}\ and\ \citenamefont
  {Boldyrev}(2010)}]{MalyshkinBoldyrev2010}%
  \BibitemOpen
  \bibfield  {author} {\bibinfo {author} {\bibfnamefont {L.~M.}\ \bibnamefont
  {Malyshkin}}\ and\ \bibinfo {author} {\bibfnamefont {S.}~\bibnamefont
  {Boldyrev}},\ }\href {\doibase 10.1103/PhysRevLett.105.215002} {\bibfield
  {journal} {\bibinfo  {journal} {\prl}\ }\textbf {\bibinfo {volume} {105}},\
  \bibinfo {pages} {215002} (\bibinfo {year} {2010})}\BibitemShut {NoStop}%
\bibitem [{\citenamefont {{Kleeorin}}\ and\ \citenamefont
  {{Rogachevskii}}(2012)}]{KleeorinRogachevskii2012}%
  \BibitemOpen
  \bibfield  {author} {\bibinfo {author} {\bibfnamefont {N.}~\bibnamefont
  {{Kleeorin}}}\ and\ \bibinfo {author} {\bibfnamefont {I.}~\bibnamefont
  {{Rogachevskii}}},\ }\href {\doibase 10.1088/0031-8949/86/01/018404}
  {\bibfield  {journal} {\bibinfo  {journal} {Physica Scripta}\ }\textbf
  {\bibinfo {volume} {86}},\ \bibinfo {pages} {018404} (\bibinfo {year}
  {2012})}\BibitemShut {NoStop}%
\bibitem [{\citenamefont {{Schober}}\ \emph
  {et~al.}(2012{\natexlab{a}})\citenamefont {{Schober}}, \citenamefont
  {{Schleicher}}, \citenamefont {{Bovino}},\ and\ \citenamefont
  {{Klessen}}}]{SchoberEtAl2012.3}%
  \BibitemOpen
  \bibfield  {author} {\bibinfo {author} {\bibfnamefont {J.}~\bibnamefont
  {{Schober}}}, \bibinfo {author} {\bibfnamefont {D.}~\bibnamefont
  {{Schleicher}}}, \bibinfo {author} {\bibfnamefont {S.}~\bibnamefont
  {{Bovino}}}, \ and\ \bibinfo {author} {\bibfnamefont {R.~S.}\ \bibnamefont
  {{Klessen}}},\ }\href {\doibase 10.1103/PhysRevE.86.066412} {\bibfield
  {journal} {\bibinfo  {journal} {\pre}\ }\textbf {\bibinfo {volume} {86}},\
  \bibinfo {eid} {066412} (\bibinfo {year} {2012}{\natexlab{a}})}\BibitemShut
  {NoStop}%
\bibitem [{\citenamefont {{Subramanian}}(1997)}]{Subramanian1997}%
  \BibitemOpen
  \bibfield  {author} {\bibinfo {author} {\bibfnamefont {K.}~\bibnamefont
  {{Subramanian}}},\ }\href@noop {} {\bibfield  {journal} {\bibinfo  {journal}
  {ArXiv e-prints: 9708216}\ } (\bibinfo {year} {1997})}\BibitemShut {NoStop}%
\bibitem [{\citenamefont {{Schekochihin}}\ \emph
  {et~al.}(2002{\natexlab{a}})\citenamefont {{Schekochihin}}, \citenamefont
  {{Boldyrev}},\ and\ \citenamefont {{Kulsrud}}}]{SchekochihinEtAl2002a}%
  \BibitemOpen
  \bibfield  {author} {\bibinfo {author} {\bibfnamefont {A.~A.}\ \bibnamefont
  {{Schekochihin}}}, \bibinfo {author} {\bibfnamefont {S.~A.}\ \bibnamefont
  {{Boldyrev}}}, \ and\ \bibinfo {author} {\bibfnamefont {R.~M.}\ \bibnamefont
  {{Kulsrud}}},\ }\href {\doibase 10.1086/338697} {\bibfield  {journal}
  {\bibinfo  {journal} {\apj}\ }\textbf {\bibinfo {volume} {567}},\ \bibinfo
  {pages} {828} (\bibinfo {year} {2002}{\natexlab{a}})}\BibitemShut {NoStop}%
\bibitem [{\citenamefont {{Schober}}\ \emph
  {et~al.}(2012{\natexlab{b}})\citenamefont {{Schober}}, \citenamefont
  {{Schleicher}}, \citenamefont {{Federrath}}, \citenamefont {{Klessen}},\ and\
  \citenamefont {{Banerjee}}}]{SchoberEtAl2012.1}%
  \BibitemOpen
  \bibfield  {author} {\bibinfo {author} {\bibfnamefont {J.}~\bibnamefont
  {{Schober}}}, \bibinfo {author} {\bibfnamefont {D.}~\bibnamefont
  {{Schleicher}}}, \bibinfo {author} {\bibfnamefont {C.}~\bibnamefont
  {{Federrath}}}, \bibinfo {author} {\bibfnamefont {R.}~\bibnamefont
  {{Klessen}}}, \ and\ \bibinfo {author} {\bibfnamefont {R.}~\bibnamefont
  {{Banerjee}}},\ }\href {\doibase 10.1103/PhysRevE.85.026303} {\bibfield
  {journal} {\bibinfo  {journal} {\pre}\ }\textbf {\bibinfo {volume} {85}},\
  \bibinfo {eid} {026303} (\bibinfo {year} {2012}{\natexlab{b}})}\BibitemShut
  {NoStop}%
\bibitem [{\citenamefont {{Kolmogorov}}(1941)}]{Kolmogorov1941}%
  \BibitemOpen
  \bibfield  {author} {\bibinfo {author} {\bibfnamefont {A.}~\bibnamefont
  {{Kolmogorov}}},\ }\href@noop {} {\bibfield  {journal} {\bibinfo  {journal}
  {Akademiia Nauk SSSR Doklady}\ }\textbf {\bibinfo {volume} {30}},\ \bibinfo
  {pages} {301} (\bibinfo {year} {1941})}\BibitemShut {NoStop}%
\bibitem [{\citenamefont {{Haugen}}\ \emph
  {et~al.}(2004{\natexlab{a}})\citenamefont {{Haugen}}, \citenamefont
  {{Brandenburg}},\ and\ \citenamefont {{Mee}}}]{HaugenEtAl2004b}%
  \BibitemOpen
  \bibfield  {author} {\bibinfo {author} {\bibfnamefont {N.~E.~L.}\
  \bibnamefont {{Haugen}}}, \bibinfo {author} {\bibfnamefont {A.}~\bibnamefont
  {{Brandenburg}}}, \ and\ \bibinfo {author} {\bibfnamefont {A.~J.}\
  \bibnamefont {{Mee}}},\ }\href {\doibase 10.1111/j.1365-2966.2004.08127.x}
  {\bibfield  {journal} {\bibinfo  {journal} {\mnras}\ }\textbf {\bibinfo
  {volume} {353}},\ \bibinfo {pages} {947} (\bibinfo {year}
  {2004}{\natexlab{a}})}\BibitemShut {NoStop}%
\bibitem [{\citenamefont {{Federrath}}\ \emph {et~al.}(2014)\citenamefont
  {{Federrath}}, \citenamefont {{Schober}}, \citenamefont {{Bovino}},\ and\
  \citenamefont {{Schleicher}}}]{FederrathEtAl2014b}%
  \BibitemOpen
  \bibfield  {author} {\bibinfo {author} {\bibfnamefont {C.}~\bibnamefont
  {{Federrath}}}, \bibinfo {author} {\bibfnamefont {J.}~\bibnamefont
  {{Schober}}}, \bibinfo {author} {\bibfnamefont {S.}~\bibnamefont {{Bovino}}},
  \ and\ \bibinfo {author} {\bibfnamefont {D.~R.~G.}\ \bibnamefont
  {{Schleicher}}},\ }\href {\doibase 10.1088/2041-8205/797/2/L19} {\bibfield
  {journal} {\bibinfo  {journal} {\apjl}\ }\textbf {\bibinfo {volume} {797}},\
  \bibinfo {eid} {L19} (\bibinfo {year} {2014})}\BibitemShut {NoStop}%
\bibitem [{\citenamefont {Burgers}(1948)}]{Burgers1948}%
  \BibitemOpen
  \bibfield  {author} {\bibinfo {author} {\bibfnamefont {J.}~\bibnamefont
  {Burgers}},\ }\href {\doibase DOI: 10.1016/S0065-2156(08)70100-5} {\emph
  {\bibinfo {title} {{A Mathematical Model Illustrating the Theory of
  Turbulence}}}},\ \bibinfo {series} {Advances in Applied Mechanics},
  Vol.~\bibinfo {volume} {1}\ (\bibinfo  {publisher} {Elsevier},\ \bibinfo
  {year} {1948})\ p.\ \bibinfo {pages} {171}\BibitemShut {NoStop}%
\bibitem [{\citenamefont {{Subramanian}}(1999)}]{Subramanian1999}%
  \BibitemOpen
  \bibfield  {author} {\bibinfo {author} {\bibfnamefont {K.}~\bibnamefont
  {{Subramanian}}},\ }\href {\doibase 10.1103/PhysRevLett.83.2957} {\bibfield
  {journal} {\bibinfo  {journal} {\prl}\ }\textbf {\bibinfo {volume} {83}},\
  \bibinfo {pages} {2957} (\bibinfo {year} {1999})}\BibitemShut {NoStop}%
\bibitem [{\citenamefont {{Schekochihin}}\ \emph
  {et~al.}(2002{\natexlab{b}})\citenamefont {{Schekochihin}}, \citenamefont
  {{Cowley}}, \citenamefont {{Hammett}}, \citenamefont {{Maron}},\ and\
  \citenamefont {{McWilliams}}}]{SchekochihinEtAl2002b}%
  \BibitemOpen
  \bibfield  {author} {\bibinfo {author} {\bibfnamefont {A.~A.}\ \bibnamefont
  {{Schekochihin}}}, \bibinfo {author} {\bibfnamefont {S.~C.}\ \bibnamefont
  {{Cowley}}}, \bibinfo {author} {\bibfnamefont {G.~W.}\ \bibnamefont
  {{Hammett}}}, \bibinfo {author} {\bibfnamefont {J.~L.}\ \bibnamefont
  {{Maron}}}, \ and\ \bibinfo {author} {\bibfnamefont {J.~C.}\ \bibnamefont
  {{McWilliams}}},\ }\href {\doibase 10.1088/1367-2630/4/1/384} {\bibfield
  {journal} {\bibinfo  {journal} {\njop}\ }\textbf {\bibinfo {volume} {4}},\
  \bibinfo {pages} {84} (\bibinfo {year} {2002}{\natexlab{b}})}\BibitemShut
  {NoStop}%
\bibitem [{\citenamefont {{Schekochihin}}\ \emph
  {et~al.}(2004{\natexlab{a}})\citenamefont {{Schekochihin}}, \citenamefont
  {{Cowley}}, \citenamefont {{Taylor}}, \citenamefont {{Hammett}},
  \citenamefont {{Maron}},\ and\ \citenamefont
  {{McWilliams}}}]{SchekochihinEtAl2004a}%
  \BibitemOpen
  \bibfield  {author} {\bibinfo {author} {\bibfnamefont {A.~A.}\ \bibnamefont
  {{Schekochihin}}}, \bibinfo {author} {\bibfnamefont {S.~C.}\ \bibnamefont
  {{Cowley}}}, \bibinfo {author} {\bibfnamefont {S.~F.}\ \bibnamefont
  {{Taylor}}}, \bibinfo {author} {\bibfnamefont {G.~W.}\ \bibnamefont
  {{Hammett}}}, \bibinfo {author} {\bibfnamefont {J.~L.}\ \bibnamefont
  {{Maron}}}, \ and\ \bibinfo {author} {\bibfnamefont {J.~C.}\ \bibnamefont
  {{McWilliams}}},\ }\href {\doibase 10.1103/PhysRevLett.92.084504} {\bibfield
  {journal} {\bibinfo  {journal} {Physical Review Letters}\ }\textbf {\bibinfo
  {volume} {92}},\ \bibinfo {eid} {084504} (\bibinfo {year}
  {2004}{\natexlab{a}})}\BibitemShut {NoStop}%
\bibitem [{\citenamefont {{Haugen}}\ \emph
  {et~al.}(2004{\natexlab{b}})\citenamefont {{Haugen}}, \citenamefont
  {{Brandenburg}},\ and\ \citenamefont {{Dobler}}}]{HaugenEtAl2004}%
  \BibitemOpen
  \bibfield  {author} {\bibinfo {author} {\bibfnamefont {N.~E.~L.}\
  \bibnamefont {{Haugen}}}, \bibinfo {author} {\bibfnamefont {A.}~\bibnamefont
  {{Brandenburg}}}, \ and\ \bibinfo {author} {\bibfnamefont {W.}~\bibnamefont
  {{Dobler}}},\ }\href {\doibase 10.1103/PhysRevE.70.016308} {\bibfield
  {journal} {\bibinfo  {journal} {\pre}\ }\textbf {\bibinfo {volume} {70}},\
  \bibinfo {pages} {016308} (\bibinfo {year} {2004}{\natexlab{b}})}\BibitemShut
  {NoStop}%
\bibitem [{\citenamefont {Federrath}\ \emph {et~al.}(2011)\citenamefont
  {Federrath}, \citenamefont {Chabrier}, \citenamefont {Schober}, \citenamefont
  {Banerjee}, \citenamefont {Klessen},\ and\ \citenamefont
  {Schleicher}}]{FederrathEtAl2011b}%
  \BibitemOpen
  \bibfield  {author} {\bibinfo {author} {\bibfnamefont {C.}~\bibnamefont
  {Federrath}}, \bibinfo {author} {\bibfnamefont {G.}~\bibnamefont {Chabrier}},
  \bibinfo {author} {\bibfnamefont {J.}~\bibnamefont {Schober}}, \bibinfo
  {author} {\bibfnamefont {R.}~\bibnamefont {Banerjee}}, \bibinfo {author}
  {\bibfnamefont {R.~S.}\ \bibnamefont {Klessen}}, \ and\ \bibinfo {author}
  {\bibfnamefont {D.~R.~G.}\ \bibnamefont {Schleicher}},\ }\href {\doibase
  10.1103/PhysRevLett.107.114504} {\bibfield  {journal} {\bibinfo  {journal}
  {\prl}\ }\textbf {\bibinfo {volume} {107}},\ \bibinfo {pages} {114504}
  (\bibinfo {year} {2011})}\BibitemShut {NoStop}%
\bibitem [{\citenamefont {{Kazantsev}}(1968)}]{Kazantsev1968}%
  \BibitemOpen
  \bibfield  {author} {\bibinfo {author} {\bibfnamefont {A.~P.}\ \bibnamefont
  {{Kazantsev}}},\ }\href@noop {} {\bibfield  {journal} {\bibinfo  {journal}
  {Soviet Journal of Experimental and Theoretical Physics}\ }\textbf {\bibinfo
  {volume} {26}},\ \bibinfo {pages} {1031} (\bibinfo {year}
  {1968})}\BibitemShut {NoStop}%
\bibitem [{\citenamefont {{Larson}}(1981)}]{Larson1981}%
  \BibitemOpen
  \bibfield  {author} {\bibinfo {author} {\bibfnamefont {R.~B.}\ \bibnamefont
  {{Larson}}},\ }\href@noop {} {\bibfield  {journal} {\bibinfo  {journal}
  {\mnras}\ }\textbf {\bibinfo {volume} {194}},\ \bibinfo {pages} {809}
  (\bibinfo {year} {1981})}\BibitemShut {NoStop}%
\bibitem [{\citenamefont {{Ossenkopf}}\ and\ \citenamefont {{Mac
  Low}}(2002)}]{OssenkopfMacLow2002}%
  \BibitemOpen
  \bibfield  {author} {\bibinfo {author} {\bibfnamefont {V.}~\bibnamefont
  {{Ossenkopf}}}\ and\ \bibinfo {author} {\bibfnamefont {M.-M.}\ \bibnamefont
  {{Mac Low}}},\ }\href {\doibase 10.1051/0004-6361:20020629} {\bibfield
  {journal} {\bibinfo  {journal} {\aap}\ }\textbf {\bibinfo {volume} {390}},\
  \bibinfo {pages} {307} (\bibinfo {year} {2002})}\BibitemShut {NoStop}%
\bibitem [{\citenamefont {{Elmegreen}}\ and\ \citenamefont
  {{Scalo}}(2004)}]{ElmegreenScalo2004}%
  \BibitemOpen
  \bibfield  {author} {\bibinfo {author} {\bibfnamefont {B.~G.}\ \bibnamefont
  {{Elmegreen}}}\ and\ \bibinfo {author} {\bibfnamefont {J.}~\bibnamefont
  {{Scalo}}},\ }\href {\doibase 10.1146/annurev.astro.41.011802.094859}
  {\bibfield  {journal} {\bibinfo  {journal} {\araa}\ }\textbf {\bibinfo
  {volume} {42}},\ \bibinfo {pages} {211} (\bibinfo {year} {2004})}\BibitemShut
  {NoStop}%
\bibitem [{\citenamefont {{Scalo}}\ and\ \citenamefont
  {{Elmegreen}}(2004)}]{ScaloElmegreen2004}%
  \BibitemOpen
  \bibfield  {author} {\bibinfo {author} {\bibfnamefont {J.}~\bibnamefont
  {{Scalo}}}\ and\ \bibinfo {author} {\bibfnamefont {B.~G.}\ \bibnamefont
  {{Elmegreen}}},\ }\href@noop {} {\bibfield  {journal} {\bibinfo  {journal}
  {\araa}\ }\textbf {\bibinfo {volume} {42}} (\bibinfo {year}
  {2004})}\BibitemShut {NoStop}%
\bibitem [{\citenamefont {{Mac Low}}\ and\ \citenamefont
  {{Klessen}}(2004)}]{MacLowKlessen2004}%
  \BibitemOpen
  \bibfield  {author} {\bibinfo {author} {\bibfnamefont {M.-M.}\ \bibnamefont
  {{Mac Low}}}\ and\ \bibinfo {author} {\bibfnamefont {R.~S.}\ \bibnamefont
  {{Klessen}}},\ }\href {\doibase 10.1103/RevModPhys.76.125} {\bibfield
  {journal} {\bibinfo  {journal} {Rev. Mod. Phys.}\ }\textbf {\bibinfo {volume}
  {76}},\ \bibinfo {pages} {125} (\bibinfo {year} {2004})}\BibitemShut
  {NoStop}%
\bibitem [{\citenamefont {{Heyer}}\ and\ \citenamefont
  {{Brunt}}(2004)}]{HeyerBrunt2004}%
  \BibitemOpen
  \bibfield  {author} {\bibinfo {author} {\bibfnamefont {M.~H.}\ \bibnamefont
  {{Heyer}}}\ and\ \bibinfo {author} {\bibfnamefont {C.~M.}\ \bibnamefont
  {{Brunt}}},\ }\href {\doibase 10.1086/425978} {\bibfield  {journal} {\bibinfo
   {journal} {\apjl}\ }\textbf {\bibinfo {volume} {615}},\ \bibinfo {pages}
  {L45} (\bibinfo {year} {2004})}\BibitemShut {NoStop}%
\bibitem [{\citenamefont {{Roman-Duval}}\ \emph {et~al.}(2011)\citenamefont
  {{Roman-Duval}}, \citenamefont {{Federrath}}, \citenamefont {{Brunt}},
  \citenamefont {{Heyer}}, \citenamefont {{Jackson}},\ and\ \citenamefont
  {{Klessen}}}]{RomanDuvalEtAl2011}%
  \BibitemOpen
  \bibfield  {author} {\bibinfo {author} {\bibfnamefont {J.}~\bibnamefont
  {{Roman-Duval}}}, \bibinfo {author} {\bibfnamefont {C.}~\bibnamefont
  {{Federrath}}}, \bibinfo {author} {\bibfnamefont {C.}~\bibnamefont
  {{Brunt}}}, \bibinfo {author} {\bibfnamefont {M.}~\bibnamefont {{Heyer}}},
  \bibinfo {author} {\bibfnamefont {J.}~\bibnamefont {{Jackson}}}, \ and\
  \bibinfo {author} {\bibfnamefont {R.~S.}\ \bibnamefont {{Klessen}}},\ }\href
  {\doibase 10.1088/0004-637X/740/2/120} {\bibfield  {journal} {\bibinfo
  {journal} {\apj}\ }\textbf {\bibinfo {volume} {740}},\ \bibinfo {eid} {120}
  (\bibinfo {year} {2011})}\BibitemShut {NoStop}%
\bibitem [{\citenamefont {{Klessen}}\ and\ \citenamefont
  {{Glover}}(2014)}]{KlessenGlover2014}%
  \BibitemOpen
  \bibfield  {author} {\bibinfo {author} {\bibfnamefont {R.~S.}\ \bibnamefont
  {{Klessen}}}\ and\ \bibinfo {author} {\bibfnamefont {S.~C.~O.}\ \bibnamefont
  {{Glover}}},\ }\href@noop {} {\bibfield  {journal} {\bibinfo  {journal}
  {ArXiv e-prints}\ } (\bibinfo {year} {2014})},\ \Eprint
  {http://arxiv.org/abs/1412.5182} {arXiv:1412.5182} \BibitemShut {NoStop}%
\bibitem [{\citenamefont {{Federrath}}(2013)}]{Federrath2013}%
  \BibitemOpen
  \bibfield  {author} {\bibinfo {author} {\bibfnamefont {C.}~\bibnamefont
  {{Federrath}}},\ }\href {\doibase 10.1093/mnras/stt1644} {\bibfield
  {journal} {\bibinfo  {journal} {\mnras}\ }\textbf {\bibinfo {volume} {436}},\
  \bibinfo {pages} {1245} (\bibinfo {year} {2013})}\BibitemShut {NoStop}%
\bibitem [{\citenamefont {{Bovino}}\ \emph {et~al.}(2013)\citenamefont
  {{Bovino}}, \citenamefont {{Schleicher}},\ and\ \citenamefont
  {{Schober}}}]{BovinoEtAl2013}%
  \BibitemOpen
  \bibfield  {author} {\bibinfo {author} {\bibfnamefont {S.}~\bibnamefont
  {{Bovino}}}, \bibinfo {author} {\bibfnamefont {D.~R.~G.}\ \bibnamefont
  {{Schleicher}}}, \ and\ \bibinfo {author} {\bibfnamefont {J.}~\bibnamefont
  {{Schober}}},\ }\href {\doibase 10.1088/1367-2630/15/1/013055} {\bibfield
  {journal} {\bibinfo  {journal} {\njop}\ }\textbf {\bibinfo {volume} {15}},\
  \bibinfo {eid} {013055} (\bibinfo {year} {2013})}\BibitemShut {NoStop}%
\bibitem [{\citenamefont {{Bhat}}\ and\ \citenamefont
  {{Subramanian}}(2014)}]{BhatSubramanian2014}%
  \BibitemOpen
  \bibfield  {author} {\bibinfo {author} {\bibfnamefont {P.}~\bibnamefont
  {{Bhat}}}\ and\ \bibinfo {author} {\bibfnamefont {K.}~\bibnamefont
  {{Subramanian}}},\ }\href {\doibase 10.1088/2041-8205/791/2/L34} {\bibfield
  {journal} {\bibinfo  {journal} {\apjl}\ }\textbf {\bibinfo {volume} {791}},\
  \bibinfo {eid} {L34} (\bibinfo {year} {2014})}\BibitemShut {NoStop}%
\bibitem [{\citenamefont {{Malyshkin}}\ and\ \citenamefont
  {{Boldyrev}}(2008)}]{MalyshkinBoldyrev2008}%
  \BibitemOpen
  \bibfield  {author} {\bibinfo {author} {\bibfnamefont {L.}~\bibnamefont
  {{Malyshkin}}}\ and\ \bibinfo {author} {\bibfnamefont {S.}~\bibnamefont
  {{Boldyrev}}},\ }\href {\doibase 10.1088/0031-8949/2008/T132/014028}
  {\bibfield  {journal} {\bibinfo  {journal} {Physica Scripta Volume T}\
  }\textbf {\bibinfo {volume} {132}},\ \bibinfo {eid} {014028} (\bibinfo {year}
  {2008})}\BibitemShut {NoStop}%
\bibitem [{\citenamefont {{Schlei\-cher}}\ \emph {et~al.}(2013)\citenamefont
  {{Schlei\-cher}}, \citenamefont {{Schober}}, \citenamefont {{Federrath}},
  \citenamefont {{Bovino}},\ and\ \citenamefont
  {{Schmidt}}}]{SchleicherEtAl2013}%
  \BibitemOpen
  \bibfield  {author} {\bibinfo {author} {\bibfnamefont {D.~R.~G.}\
  \bibnamefont {{Schlei\-cher}}}, \bibinfo {author} {\bibfnamefont
  {J.}~\bibnamefont {{Schober}}}, \bibinfo {author} {\bibfnamefont
  {C.}~\bibnamefont {{Federrath}}}, \bibinfo {author} {\bibfnamefont
  {S.}~\bibnamefont {{Bovino}}}, \ and\ \bibinfo {author} {\bibfnamefont
  {W.}~\bibnamefont {{Schmidt}}},\ }\href {\doibase
  10.1088/1367-2630/15/2/023017} {\bibfield  {journal} {\bibinfo  {journal}
  {\njop}\ }\textbf {\bibinfo {volume} {15}},\ \bibinfo {pages} {023017}
  (\bibinfo {year} {2013})}\BibitemShut {NoStop}%
\bibitem [{\citenamefont {{Kulsrud}}\ and\ \citenamefont
  {{Anderson}}(1992)}]{KulsrudAnderson1992}%
  \BibitemOpen
  \bibfield  {author} {\bibinfo {author} {\bibfnamefont {R.~M.}\ \bibnamefont
  {{Kulsrud}}}\ and\ \bibinfo {author} {\bibfnamefont {S.~W.}\ \bibnamefont
  {{Anderson}}},\ }\href {\doibase 10.1086/171743} {\bibfield  {journal}
  {\bibinfo  {journal} {\apj}\ }\textbf {\bibinfo {volume} {396}},\ \bibinfo
  {pages} {606} (\bibinfo {year} {1992})}\BibitemShut {NoStop}%
\bibitem [{\citenamefont {{Maron}}\ and\ \citenamefont
  {{Cowley}}(2001)}]{MaronCowley2001}%
  \BibitemOpen
  \bibfield  {author} {\bibinfo {author} {\bibfnamefont {J.}~\bibnamefont
  {{Maron}}}\ and\ \bibinfo {author} {\bibfnamefont {S.}~\bibnamefont
  {{Cowley}}},\ }\href@noop {} {\bibfield  {journal} {\bibinfo  {journal}
  {ArXiv Astrophysics e-prints}\ } (\bibinfo {year} {2001})},\ \Eprint
  {http://arxiv.org/abs/astro-ph/0111008} {astro-ph/0111008} \BibitemShut
  {NoStop}%
\bibitem [{\citenamefont {{She}}\ and\ \citenamefont
  {{Leveque}}(1994)}]{SheLeveque1994}%
  \BibitemOpen
  \bibfield  {author} {\bibinfo {author} {\bibfnamefont {Z.-S.}\ \bibnamefont
  {{She}}}\ and\ \bibinfo {author} {\bibfnamefont {E.}~\bibnamefont
  {{Leveque}}},\ }\href {\doibase 10.1103/PhysRevLett.72.336} {\bibfield
  {journal} {\bibinfo  {journal} {\prl}\ }\textbf {\bibinfo {volume} {72}},\
  \bibinfo {pages} {336} (\bibinfo {year} {1994})}\BibitemShut {NoStop}%
\bibitem [{\citenamefont {{Boldyrev}}\ \emph {et~al.}(2002)\citenamefont
  {{Boldyrev}}, \citenamefont {{Nordlund}},\ and\ \citenamefont
  {{Padoan}}}]{Boldyrev2002}%
  \BibitemOpen
  \bibfield  {author} {\bibinfo {author} {\bibfnamefont {S.}~\bibnamefont
  {{Boldyrev}}}, \bibinfo {author} {\bibfnamefont {{\AA}.}~\bibnamefont
  {{Nordlund}}}, \ and\ \bibinfo {author} {\bibfnamefont {P.}~\bibnamefont
  {{Padoan}}},\ }\href {\doibase 10.1086/340758} {\bibfield  {journal}
  {\bibinfo  {journal} {\apj}\ }\textbf {\bibinfo {volume} {573}},\ \bibinfo
  {pages} {678} (\bibinfo {year} {2002})}\BibitemShut {NoStop}%
\bibitem [{\citenamefont {{Federrath}}\ \emph {et~al.}(2010)\citenamefont
  {{Federrath}}, \citenamefont {{Roman-Duval}}, \citenamefont {{Klessen}},
  \citenamefont {{Schmidt}},\ and\ \citenamefont {{Mac
  Low}}}]{FederrathEtAl2010}%
  \BibitemOpen
  \bibfield  {author} {\bibinfo {author} {\bibfnamefont {C.}~\bibnamefont
  {{Federrath}}}, \bibinfo {author} {\bibfnamefont {J.}~\bibnamefont
  {{Roman-Duval}}}, \bibinfo {author} {\bibfnamefont {R.~S.}\ \bibnamefont
  {{Klessen}}}, \bibinfo {author} {\bibfnamefont {W.}~\bibnamefont
  {{Schmidt}}}, \ and\ \bibinfo {author} {\bibfnamefont {M.-M.}\ \bibnamefont
  {{Mac Low}}},\ }\href {\doibase 10.1051/0004-6361/200912437} {\bibfield
  {journal} {\bibinfo  {journal} {\aap}\ }\textbf {\bibinfo {volume} {512}},\
  \bibinfo {eid} {A81} (\bibinfo {year} {2010})}\BibitemShut {NoStop}%
\bibitem [{\citenamefont {{Meneguzzi}}\ \emph {et~al.}(1981)\citenamefont
  {{Meneguzzi}}, \citenamefont {{Frisch}},\ and\ \citenamefont
  {{Pouquet}}}]{MeneguzziEtAl1981}%
  \BibitemOpen
  \bibfield  {author} {\bibinfo {author} {\bibfnamefont {M.}~\bibnamefont
  {{Meneguzzi}}}, \bibinfo {author} {\bibfnamefont {U.}~\bibnamefont
  {{Frisch}}}, \ and\ \bibinfo {author} {\bibfnamefont {A.}~\bibnamefont
  {{Pouquet}}},\ }\href {\doibase 10.1103/PhysRevLett.47.1060} {\bibfield
  {journal} {\bibinfo  {journal} {\prl}\ }\textbf {\bibinfo {volume} {47}},\
  \bibinfo {pages} {1060} (\bibinfo {year} {1981})}\BibitemShut {NoStop}%
\bibitem [{\citenamefont {{Schekochihin}}\ \emph
  {et~al.}(2004{\natexlab{b}})\citenamefont {{Schekochihin}}, \citenamefont
  {{Cowley}}, \citenamefont {{Taylor}}, \citenamefont {{Maron}},\ and\
  \citenamefont {{McWilliams}}}]{SchekochihinEtAl2004.2}%
  \BibitemOpen
  \bibfield  {author} {\bibinfo {author} {\bibfnamefont {A.~A.}\ \bibnamefont
  {{Schekochihin}}}, \bibinfo {author} {\bibfnamefont {S.~C.}\ \bibnamefont
  {{Cowley}}}, \bibinfo {author} {\bibfnamefont {S.~F.}\ \bibnamefont
  {{Taylor}}}, \bibinfo {author} {\bibfnamefont {J.~L.}\ \bibnamefont
  {{Maron}}}, \ and\ \bibinfo {author} {\bibfnamefont {J.~C.}\ \bibnamefont
  {{McWilliams}}},\ }\href {\doibase 10.1086/422547} {\bibfield  {journal}
  {\bibinfo  {journal} {\apj}\ }\textbf {\bibinfo {volume} {612}},\ \bibinfo
  {pages} {276} (\bibinfo {year} {2004}{\natexlab{b}})}\BibitemShut {NoStop}%
\bibitem [{\citenamefont {{Grete}}\ \emph {et~al.}(2015)\citenamefont
  {{Grete}}, \citenamefont {{Vlaykov}}, \citenamefont {{Schmidt}},
  \citenamefont {{Schleicher}},\ and\ \citenamefont
  {{Federrath}}}]{GreteEtAl2015}%
  \BibitemOpen
  \bibfield  {author} {\bibinfo {author} {\bibfnamefont {P.}~\bibnamefont
  {{Grete}}}, \bibinfo {author} {\bibfnamefont {D.~G.}\ \bibnamefont
  {{Vlaykov}}}, \bibinfo {author} {\bibfnamefont {W.}~\bibnamefont
  {{Schmidt}}}, \bibinfo {author} {\bibfnamefont {D.~R.~G.}\ \bibnamefont
  {{Schleicher}}}, \ and\ \bibinfo {author} {\bibfnamefont {C.}~\bibnamefont
  {{Federrath}}},\ }\href {\doibase 10.1088/1367-2630/17/2/023070} {\bibfield
  {journal} {\bibinfo  {journal} {\njop}\ }\textbf {\bibinfo {volume} {17}},\
  \bibinfo {eid} {023070} (\bibinfo {year} {2015})}\BibitemShut {NoStop}%
\bibitem [{\citenamefont {{Brandenburg}}\ \emph {et~al.}(1996)\citenamefont
  {{Brandenburg}}, \citenamefont {{Enqvist}},\ and\ \citenamefont
  {{Olesen}}}]{BrandenburgEnqvistOlesen1996}%
  \BibitemOpen
  \bibfield  {author} {\bibinfo {author} {\bibfnamefont {A.}~\bibnamefont
  {{Brandenburg}}}, \bibinfo {author} {\bibfnamefont {K.}~\bibnamefont
  {{Enqvist}}}, \ and\ \bibinfo {author} {\bibfnamefont {P.}~\bibnamefont
  {{Olesen}}},\ }\href {\doibase 10.1103/PhysRevD.54.1291} {\bibfield
  {journal} {\bibinfo  {journal} {\prd}\ }\textbf {\bibinfo {volume} {54}},\
  \bibinfo {pages} {1291} (\bibinfo {year} {1996})}\BibitemShut {NoStop}%
\bibitem [{\citenamefont {{Plunian}}\ \emph {et~al.}(2013)\citenamefont
  {{Plunian}}, \citenamefont {{Stepanov}},\ and\ \citenamefont
  {{Frick}}}]{PlunianStepanovFrick2013}%
  \BibitemOpen
  \bibfield  {author} {\bibinfo {author} {\bibfnamefont {F.}~\bibnamefont
  {{Plunian}}}, \bibinfo {author} {\bibfnamefont {R.}~\bibnamefont
  {{Stepanov}}}, \ and\ \bibinfo {author} {\bibfnamefont {P.}~\bibnamefont
  {{Frick}}},\ }\href {\doibase 10.1016/j.physrep.2012.09.001} {\bibfield
  {journal} {\bibinfo  {journal} {\physrep}\ }\textbf {\bibinfo {volume}
  {523}},\ \bibinfo {pages} {1} (\bibinfo {year} {2013})}\BibitemShut {NoStop}%
\bibitem [{\citenamefont {{Stepanov}}\ and\ \citenamefont
  {{Plunian}}(2008)}]{StepanovPlunian2008}%
  \BibitemOpen
  \bibfield  {author} {\bibinfo {author} {\bibfnamefont {R.}~\bibnamefont
  {{Stepanov}}}\ and\ \bibinfo {author} {\bibfnamefont {F.}~\bibnamefont
  {{Plunian}}},\ }\href {\doibase 10.1086/587795} {\bibfield  {journal}
  {\bibinfo  {journal} {\apj}\ }\textbf {\bibinfo {volume} {680}},\ \bibinfo
  {pages} {809} (\bibinfo {year} {2008})}\BibitemShut {NoStop}%
\end{thebibliography}
\end{document}